\documentclass[fleqn,usenatbib]{mnras}

\usepackage{newtxtext,newtxmath}

\usepackage[T1]{fontenc}

\DeclareRobustCommand{\VAN}[3]{#2}
\let\VANthebibliography\thebibliography
\def\thebibliography{\DeclareRobustCommand{\VAN}[3]{##3}\VANthebibliography}

\usepackage{graphicx}	
\usepackage{amsmath}	

\usepackage{newtxmath}
\usepackage[flushleft,referable]{threeparttablex}
\usepackage{longtable}
\usepackage{natbib}
\usepackage{enumitem}

\newcommand{\LyA}{Ly$\alpha$}
\newcommand{\SSFR}{$\Sigma_{\rm SFR}$}
\newcommand{\sSSFR}{$\Sigma_{\rm sSFR}$}
\newcommand{\Vout}{$v_{\rm{out}}$}
\newcommand{\Vlis}{$\Delta v_{\rm{LIS}}$}
\newcommand{\Vfrc}{$v_{\rm{80}}$}
\newcommand{\Vmax}{$v_{\rm{max}}$}
\newcommand{\Vlya}{$\Delta v_{\rm{Ly\alpha}}$}
\newcommand{\RE}{$R_{\rm{E}}$}
\newcommand{\HA}{$\rm{H}\alpha$}
\newcommand{\HB}{$\rm{H}\beta$}
\newcommand{\Mste}{$M_{\star}$}
\newcommand{\Mdyn}{$M_{\rm{dyn}}$}
\newcommand{\Mgas}{$M_{\rm{gas}}$}
\newcommand{\Mbar}{$M_{\rm{bar}}$}
\newcommand{\Wlis}{$W_{\rm{LIS}}$}
\newcommand{\Wlya}{$W_{\rm{Ly}\alpha}$}
\newcommand{\depth}{$\tau_{a}$}

\newcommand{\aref}[1]{\hyperref[#1]{Appendix~\ref{#1}}}

\DeclareUnicodeCharacter{2212}{-}

\title[MOSDEF-LRIS: Outflows and Galaxy Properties]{The MOSDEF-LRIS Survey: Connection between Galactic-scale Outflows and the Properties of z $\sim$ 2 Star-forming Galaxies}
\author[A. Weldon et al.]{
Andrew Weldon,$^{1}$\thanks{E-mail: aweld004@ucr.edu}
Naveen A. Reddy,$^{1}$
Michael W. Topping,$^{2}$
Alice E. Shapley,$^{3}$
Ryan L. Sanders,$^{4, 5}$
\newauthor
Xinnan Du,$^{6}$
Sedona H. Price,$^{7}$
Alison L. Coil,$^{8}$
Brian Siana,$^{1}$
Bahram Mobasher,$^{1}$
Tara Fetherolf,$^{1,9}$
\newauthor
Irene Shivaei,$^{2,5}$ and
Saeed Rezaee$^{1}$
\\
$^{1}$Department of Physics and Astronomy, University of California, Riverside, 900 University Avenue, Riverside, CA 92521, USA\\
$^{2}$Steward Observatory, University of Arizona, 933 N Cherry Ave, Tucson, AZ 85721, USA\\
$^{3}$Physics \& Astronomy Department, University of California: Los Angeles, 430 Portola Plaza, Los Angeles, CA 90095, USA\\
$^{4}$Department of Physics and Astronomy, University of California, Davis, One Shields Ave, Davis, CA 95616, USA\\
$^{5}$Hubble Fellow\\
$^{6}$Kavli Institute for Particle Astrophysics \& Cosmology, P. O. Box 2450, Stanford University, Stanford, CA 94305, USA\\
$^{7}$Max-Planck-Institut für extraterrestrische Physik (MPE), Giessenbachstr. 1, D-85748 Garching, Germany\\
$^{8}$Center for Astrophysics and Space Sciences, Department of Physics, University of California, San Diego, 9500 Gilman Dr., La Jolla, CA 92093-0424, USA\\
$^{9}$UC Chancellor’s Fellow
}

\date{Accepted XXX. Received YYY; in original form ZZZ}

\pubyear{2022}

\begin{document}
\label{firstpage}
\pagerange{\pageref{firstpage}--\pageref{lastpage}}
\maketitle

\begin{abstract}

We investigate the conditions that facilitate galactic-scale outflows using a sample of 155 typical star-forming galaxies at $z$ $\sim$ 2 drawn from the MOSFIRE Deep Evolution Field (MOSDEF) survey. The sample includes deep rest-frame UV spectroscopy from the Keck Low-Resolution Imaging Spectrometer (LRIS), which provides spectral coverage of several low-ionisation interstellar (LIS) metal absorption lines and \LyA\ emission.  Outflow velocities are calculated from the centroids of the LIS absorption and/or \LyA\ emission, as well as the highest-velocity component of the outflow from the blue wings of the LIS absorption lines. Outflow velocities are found to be marginally correlated or independent of galaxy properties, such as star-formation rate (SFR) and star-formation rate surface density (\SSFR). Outflow velocity scales with SFR as a power-law with index 0.24, which suggests that the outflows may be primarily driven by mechanical energy generated by supernovae explosions, as opposed to radiation pressure acting on dusty material. On the other hand, outflow velocity and \SSFR\ are not significantly correlated, which may be due to the limited dynamic range of \SSFR\ probed by our sample. The relationship between outflow velocity and \SSFR\ normalised by stellar mass (\sSSFR), as a proxy for gravitational potential, suggests that strong outflows (e.g., > 200 km s$^{-1}$) appear ubiquitous above a threshold of log(\sSSFR/$\rm{yr}^{-1}\ \rm{kpc}^{-2}$) $\sim$ $-$11.3, and that above this threshold, outflow velocity uncouples from \sSSFR. These results highlight the need for higher resolution spectroscopic data and spatially resolved imaging to test the driving mechanisms of outflows predicted by theory.

\end{abstract}




\section{Introduction}

The evolution of galaxies is influenced by the flow of baryons. Galaxies accrete cold gas from filaments in the cosmic web, convert the gas into stars, and eject metal-enriched gas from the interstellar medium (ISM) into the circumgalactic medium (CGM) or possibly beyond into the intergalactic medium (IGM). One important component of this cycle is galactic-scale outflows, which enriches the CGM and IGM with metals, modulates the metallicity within galaxies  \citep[e.g.,][]{Tremonti04, Dalcanton07, Finlator08}, and depletes the availability of cold gas causing a suppression of star formation \citep[e.g.,][]{Scannapieco05, Croton06}. Outflows also appear to be an important factor in the creation of low-column-density channels in the ISM, allowing for the escape of ionising photons \citep[e.g.,][]{Gnedin08, Leitet13, Ma16, Reddy16, Gazagnes18, Reddy22}.

Galactic outflows are a common feature of actively star-forming galaxies, with observations of such flows in local galaxies \citep[e.g.,][]{Heckman00, Chen10, Roberts20} and high-redshift galaxies \citep[e.g.,][]{Shapley03, Steidel10, Davies19}. However, the physical mechanisms that generate and sustain outflows remain an open question. In star-forming galaxies, outflows are theorised to be driven by energy injected into the ISM by supernovae; radiation pressure acting on cool, dusty material; or a combination of these mechanisms \citep{Chevalier85, Murray05, Murray11}. Outflow velocity should then scale with star formation properties as the level of star formation activity, which peaks at $z$ $\sim$ 1 -- 3 \citep{Madau14}, sets the effectiveness of these mechanisms.

Galactic outflows are often probed by either blueshifted interstellar absorption or redshifted resonantly-scattered emission. Observations of blueshifted interstellar absorption lines for both local and high-redshift galaxies have found that outflow velocity (\Vout) increases with several galactic properties such as stellar mass, star-formation rate (SFR), and star-formation-rate surface density (\SSFR) \citep[e.g.,][]{Sato09, Weiner09, Chen10, Rubin10, Steidel10, Law12, Martin12, Rubin14, Chisholm15, Heckman15, Bordoloi16, Davies19, Prusinski21}. However, the existence of a \Vout$-$SFR or \Vout$-$\SSFR\ relation is still debated. There is general agreement at low redshifts ($z$ $\lesssim$ 0.3) that the relations are significant, weak power laws \citep[\Vout\ $\propto$ SFR$^{0.15-0.35}$;][\Vout\ $\propto$ \SSFR$^{0.1}$; \citealt{Chen10}]{Martin05, Chisholm15, Sugahara17}. At higher redshifts (1 $\lesssim$ $z$ $\lesssim$ 2) there is disagreement on the significance of the \Vout$-$\SSFR\ relation \citep{Steidel10, Kornei12, Law12, Rubin14, Prusinski21}. This tension may be due to the use of ions (e.g., C IV, Si IV, Si II, and Mg II) that trace different components of the outflowing gas, different methods to parameterise outflow kinematics, and differences in the methodology used to estimate \SSFR\ \citep[see discussions in][]{Ho16, Heckman17}.

Rest-UV spectra of $z$ $\sim$ 2 star forming galaxies contain a wealth of emission and low-ionisation interstellar (LIS) absorption metal lines (e.g., Si II, O I, C II, and Fe II) of cool, diffuse interstellar gas transported by galactic-scale outflows. Due to the difficulty of obtaining rest-UV spectra with sufficient S/N for typically faint high-redshift galaxies, previous studies have been primarily limited to gravitationally-lensed Lyman Break galaxies (LBGs), very luminous LBGs, or high S/N composite spectra to infer average outflow properties \citep{Pettini02, Shapley03, Rubin10, Jones12, Bordoloi16, Du18}. However, recent studies using deep observations have been able to focus on individual galaxies at $z$ $\sim$ 2 \citep{Forster19, Davies20}. Here, we use a sample of 155 galaxies drawn from the MOSFIRE Deep Evolution Field \citep[MOSDEF;][]{MOSDEF} Survey with additional deep ($\sim$7.5 hrs) rest-UV observations from the Keck Low Resolution Imaging Spectrometer \citep[LRIS;][]{LRIS, Steidel03}. The combination of rest-optical spectra from the MOSFIRE near-IR spectrograph and rest-UV spectra from LRIS creates an ideal dataset for investigating the relationship between outflows and the physical properties of the host galaxy at high redshift on an individual galaxy basis. The primary objectives of this study are to (1) explore which, if any, galactic properties correlate with outflow properties; and (2) determine the primary driving mechanisms of outflows in $z$ $\sim$ 2 -- 3 galaxies.

The outline of this paper is as follows. In Section \ref{sec:Data}, we introduce the sample and stellar population models used in this work. Section \ref{sec:3} describes the approach for estimating SFR, \SSFR, outflow velocity, and other galaxy properties. In Section \ref{sec:4}, we present our main results on the correlations between outflow velocities and measured galaxy properties. We discuss the physical context behind these results in Section \ref{sec:5} and summarise our conclusions in Section \ref{sec:con}. Throughout this paper, we adopt a standard cosmology with $\Omega_{\Lambda}$ = 0.7, $\Omega_{\rm{M}}$ = 0.3, and $\rm{H}_{0}$ = 70 km $\rm{s}^{-1} \rm{ Mpc}^{-1}$. All wavelengths are presented in the vacuum frame.

\section{Data}
\label{sec:Data}

\subsection{MOSDEF Spectroscopy}

Our analysis utilises rest-optical spectra from the MOSDEF Survey which targeted $\approx$1500 $H$-band selected galaxies in the CANDELS fields \citep{Grogin11, Koekemoer11}. The survey used the MOSFIRE spectrograph \citep{McLean12} on the 10m Keck I telescope over 48.5 nights from 2012 -- 2016 to obtain moderate-resolution ($R$ $\sim$ 3000−-3600) near-infrared spectra. Galaxies were targeted for spectroscopy based on pre-existing spectroscopic, grism, or photometric redshifts that placed them into one of three redshift ranges ($z$ = 1.37 −- 1.70, $z$ = 2.09 -− 2.61, and $z$ = 2.95 −- 3.80) where strong rest-frame optical emission lines lie in the $YJHK$ transmission windows. For full details regarding the survey (targeting, data reduction, and sample properties), we refer readers to \cite{MOSDEF}.

Emission line fluxes were measured from the MOSFIRE spectra by simultaneously fitting a line with a Gaussian function and a linear continuum component. Two Gaussian functions were used to fit the [O II] doublet, while three were used to fit \HA\ and the [N II] doublet. Systemic redshifts were derived from the strongest emission line, usually \HA\ or [O III]$\lambda$5008, and were used to fit the other rest-frame optical nebular emission lines. Line fluxes and errors were derived by perturbing the spectra by its error spectrum to generate 1000 realisations, remeasuring the line fluxes from these realisations, and calculating the average line fluxes and their dispersion from the realisations. Further details on emission line measurements and slit loss corrections are given in \cite{MOSDEF} and \cite{Reddy15}.

Galaxy sizes and inclinations were estimated from the effective radius (\RE), within which half the total light of the galaxy is contained, and the axis ratio ($b$/$a$), respectively, measured by \cite{vanderWel14}\footnote{\url{https://users.ugent.be/~avdrwel/research.html}} using GALFIT \citep{Peng10} on HST/F160W images from the CANDELS survey. 

\subsection{MOSDEF-LRIS Spectroscopy}

\begin{figure*}
  \includegraphics[width=\linewidth, keepaspectratio]{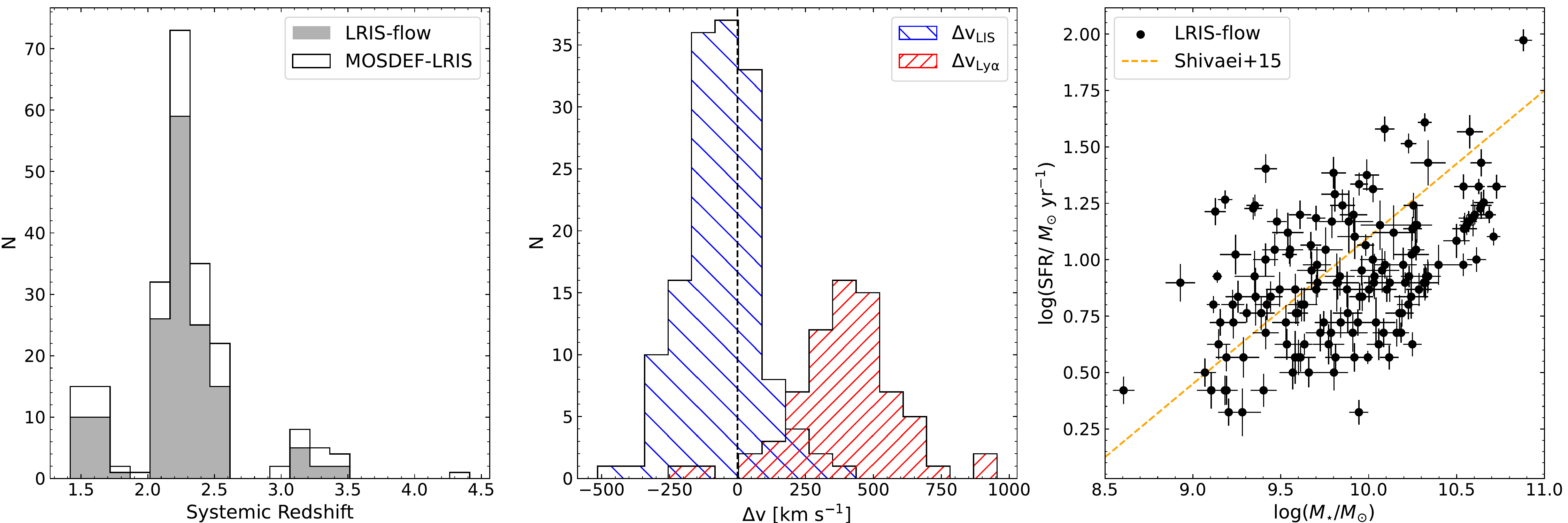}
  \vspace{-0.5cm}
  \caption{Properties of the LRIS-flow sample. \textit{Left}: Redshift distribution. Open black and solid grey histograms represent the MOSDEF-LRIS sample with systemic redshift measurements (215 objects) and the LRIS-flow sample with additional LIS or \LyA\ redshifts (155 objects), respectively. \textit{Center}: Velocity distribution. Blue and red hashed histograms denote the distribution of centroid velocities of the LIS absorption lines (149 objects) and \LyA\ emission (72 objects), respectively. \textit{Right}: SFR versus stellar mass. SFRs and stellar masses are derived from SED modelling (Section \ref{sec:sample}). For comparison, the SFR--stellar mass relation derived in \protect\cite{Shivaei15} the parent MOSDEF sample is shown as an orange dash line. The SFRs in the study are derived from dust-corrected \HA\ fluxes.} 
  \label{fig:sample-3}
\end{figure*}

In this study, we use a sample of 155 galaxies drawn from the MOSDEF survey with follow-up rest-UV frame LRIS observations. Here, we briefly summarise the sample and refer readers to \cite{topping2020} for more details. From the MOSDEF Survey, objects were prioritised for LRIS spectroscopy based on detections of rest-optical emission lines (\HB, [OIII], \HA, and [NII]). Higher priority was given to objects with confirmed spectroscopic redshifts at 1.90 $\leq$ $z$ $\leq$ 2.65. Additional objects were selected in the following order of priority: objects with \HA, \HB, and [OIII] detected at $\geq$ 3$\sigma$ and an upper limit on [NII]; objects with a confirmed systemic redshift from MOSDEF; objects observed as part of the MOSDEF survey without a successful systemic redshift measurement, but with a prior spectroscopic or photometric redshift; and finally, objects not observed with MOSFIRE, but with a prior redshift from the 3D-HST survey that placed them within the redshift ranges and magnitude limit of the MOSDEF survey. In total, 260 galaxies were selected for follow-up observations with LRIS\footnote{Systemic redshift measured by MOSDEF were obtained for 214 galaxies, while the remaining 45 galaxies either have a spectroscopic redshift prior to the MOSDEF survey or a photometric redshift.}.

LRIS observations were obtained over nine nights in 2017 and 2018 in the COSMOS, GOODS-S, GOODS-N, and AEGIS fields using nine multi-object slit masks milled with 1\farcs2 slits. The instrumental setup included a dichroic to split the incoming beam at $\sim$5000\AA\ into the blue and red arms of LRIS. We configured the blue side with the 400 lines/mm grism, and the red side with the 600 lines/mm grating. This configuration provided continuous spectral coverage from the atmospheric cut-off at 3100\AA\ up to a typical wavelength of $\sim$7000\AA, depending on the position of the slit within the spectroscopic field of view. The seeing ranged from 0\farcs6 to 1\farcs2 with a typical value of 0\farcs8. The process to create the 1D spectra used in this study is described in \cite{topping2020} and we refer readers there for more details. The rest-frame spectra were continuum normalised around each LIS absorption line. The local continuum was determined by fitting a linear model between the average flux in two spectral windows, bluewards and redwards of the LIS absorption line. The spectral windows, listed in Table \ref{tab:spectral_windows}, were chosen to bracket the line and free of other spectral features.

\subsection{Redshift Measurements}

Due to large-scale galaxy outflows, low-ionisation interstellar absorption lines and the \LyA\ emission are Doppler shifted away from the MOSDEF systemic redshift ($z_{\rm{sys}}$) measured from strong rest-frame optical emission lines. LIS absorption line ($z_{\rm{LIS}}$) and \LyA\ ($z_{\rm{Ly\alpha}}$) redshifts and uncertainties were measured using the procedures described in \cite{topping2020}.  Briefly, these redshifts were obtained by fitting lines with Gaussian functions and a quadratic function for the local continuum and calculating the centroids of the Gaussians.  Redshift uncertainties were determined by perturbing the LRIS spectra by the corresponding error spectra, refitting lines, and recalculating the centroids. Any LIS absorption lines with poor fits were excluded in the calculation of the average $z_{\rm{LIS}}$, typically based on two lines, for a given galaxy.

\subsection{Sample Selection}
\label{sec:sample}

For our analysis of outflows, 167 galaxies were initially selected with $z_{\rm{sys}}$ determined from MOSDEF observations, $z_{\rm{LIS}}$ and/or $z_{\rm{Ly\alpha}}$ measurements from LRIS observations. Additionally, as we are interested in star-forming galaxies, 12 galaxies were removed for possible AGNs based based on IR emission, X-ray emission, and the [NII]/\HA\ line ratio \citep{Coil15, Azadi17, Azadi18, Leung19}. These requirements reduced the sample of 260 MOSDEF-LRIS galaxies to 155 galaxies (hereafter the “LRIS-flow” sample). For nine galaxies, the MOSDEF or LRIS slit includes close, unresolved galaxies such that there may be a mismatch between the measured redshifts. These galaxies are included in the LRIS-flow sample, and we note that our final results are not affected by their inclusion. As shown in Figure \ref{fig:sample-3}, the LRIS-flow sample has a redshift range of 1.42 $\leq$ $z$ $\leq$ 3.48 with a median redshift of 2.24, centroid LIS absorption line velocities (\Vlis) from $-$510 km s$^{-1}$ to 380 km s$^{-1}$ with a mean of $-$60$\pm$10 km s$^{-1}$, and centroid \LyA\ emission velocities (\Vlya) from $-$190 km s$^{-1}$ to 950 km s$^{-1}$ with a mean of 400$\pm$23 km s$^{-1}$.

\begin{table}
  \centering
  \caption{Spectral Windows}
  \label{tab:spectral_windows}
  \begin{threeparttable}
    \begin{tabular}{lccc}
        \hline
        \hline
        Line  & $\lambda$ (\AA)\tnote{a} & Blue Window (\AA)\tnote{b} & Red Window (\AA)\tnote{b}\\
        \hline
   
        Ly$\alpha$ & 1215.67 & 1195 - 1202 & 1225 - 1235\\
        SiII       & 1260.42 & 1245 - 1252 & 1270 - 1275\\
        OI         & 1302.17 & 1285 - 1293 & 1312 - 1318\\
        CII        & 1334.53 & 1320 - 1330 & 1342 - 1351\\
        SiII       & 1526.71 & 1512 - 1520 & 1535 - 1540\\
        FeII       & 1608.45 & 1590 - 1600 & 1616 - 1620\\
    \hline
    \end{tabular}
    \begin{tablenotes}
        \item[a] Rest-frame vacuum wavelength, taken from the Atomic Spectra Database website of the National Institute of Standards and Technology (NIST), https://www.nist.gov/pml/atomic-spectra-database.
        \item[b] Wavelength window over which continuum fitting was performed.
    \end{tablenotes}

  \end{threeparttable}
\end{table}

\section{Measurements}
\label{sec:3}

\subsection{Stellar Population Properties}
\label{sec:SED}

Stellar masses (\Mste) and SFRs of the LRIS-flow sample were derived from spectral energy distribution (SED) modelling. Here, we briefly describe the models used and refer readers to \cite{Reddy15} for more details. The models were created adopting a \citet[hereafter BC03]{BC03} stellar population synthesis model, \cite{Chabrier03} initial mass function, constant star formation histories (SFH), Small Magellanic Cloud (SMC) attenuation curve \citep{Fitzpatrick90, Gordon03}, and sub-solar metallicity $Z$ = 0.28$Z_{\odot}$\footnote{Recent MOSDEF studies have suggested that a \cite{Calzetti00} attenuation curve and solar metallicities provide a better description for high-mass star-forming galaxies at z $\sim$ 2, compared to an SMC  attenuation curve  with sub-solar metallicities \citep{Reddy18a, Shivaei20}. If instead we assume a \cite{Calzetti00} attenuation curve and solar metallicities for high-mass galaxies, on average, stellar masses are lower and SFRs are higher. However, our main results do not significantly change if we were to alter the assumed attenuation curve.}. A lower age limit of 50 Myr was imposed, based on the typical dynamical timescale of $z$ $\sim$ 2 galaxies \citep{Reddy12}. The combination of the steeper SMC attenuation curve, which has been found to best reproduce the dust obscurations of typical star-forming galaxies at $z$ $\sim$ 2 based on far-infrared data \citep{Reddy18a}, and sub-solar metallicity provide self-consistent SFRs with those derived using other methods \citep{Reddy18b, Theios19}\footnote{See also Appendix A of \cite{Reddy22}.}. The best-fit stellar population parameters and their errors were obtained by perturbing the photometry, refitting the models, and taking the median and dispersion in the resulting parameters, respectively. As shown in the right panel of Figure \ref{fig:sample-3}, the LRIS-flow sample covers a range of typical star-forming galaxies. The LRIS-flow sample has a stellar mass range of 8.61 < log($M_{\star}$/$M_{\odot}$) < 10.88 with a median log($M_{\star}$/$M_{\odot}$) of 9.89 and SFR range from 0.32 < log(SFR/$M_{\odot}$ yr$^{-1}$) < 1.97 with a median log(SFR/$M_{\odot}$ yr$^{-1}$) of 0.9.

\subsection{SFR and SFR Surface Densities}
\label{sec:SSFR_clac}
We calculate \HA\ SFRs (SFR[\HA]) from \HA\ and \HB\ flux measurements corrected for dust using the Balmer decrement. Following the methodology presented in \cite{Reddy15},  \HA\ luminosities are corrected for attenuation assuming a \cite{Cardelli89} Galactic extinction curve\footnote{\cite{Reddy20} found that the nebular attenuation curve is similar in shape to that of the Galactic extinction curve \citep{Cardelli89}.} and converted to SFRs using the conversion factor from \cite{Reddy18b}, $\rm{3.236\times10^{−42}}$ $M_{\odot}$ yr$^{-1}$ ergs$^{-1}$ s, for BC03 stellar population synthesis model and sub-solar metallicity adopted for the SED fitting (see Section \ref{sec:SED}). SFR[\HA] is calculated for objects with significant detections (S/N > 3) of \HA\ and \HB. For objects where \HB\ is undetected, 3$\sigma$ upper limits are assigned. As the SFR from SED fitting is tightly correlated with stellar mass (i.e., both quantities are sensitive to the normalisation of the best-fit SED), we have chosen to focus on SFR[SED] when discussing SFR and SFR[\HA] when discussing the specific star-formation rate (sSFR = SFR/$M_{\star}$). As discussed in previous studies, there is a general agreement between SFR[SED] and SFR[\HA] for MOSDEF galaxies \citep[e.g.,][]{Reddy15, Shivaei16, Azadi18, Reddy22}.

Along with the star-formation rate, the mechanisms that drive outflows may be enhanced in regions of compact star formation, so we define the star-formation-rate surface density (\SSFR) as
\begin{equation}
	   \Sigma_{\rm{SFR}} = \frac{\rm{SFR[SED]}}{2\pi R_{\rm{E}}^{2}}.
\end{equation}
At a given \SSFR, outflows may be more effectively launched from a shallow galaxy potential (i.e., low stellar mass) relative to a deep potential.
To examine the dependence of outflow velocity on both \SSFR\ and the galaxy potential, we compute the specific star-formation-rate surface density (\sSSFR):
\begin{equation}
	   \Sigma_{\rm{SFR}}/M_{X} = \frac{\Sigma_{\rm{SFR[H\alpha]}}}{M_{X}} = \frac{\rm{SFR[H\alpha]}}{2\pi R_{\rm{E}}^{2}M_{X}},
	   \label{equ:sSSFR}
\end{equation}
where $M_{X}$ can be stellar, dynamical, or baryonic mass (Section \ref{sec:alt_mass}). As discussed in \cite{Price20}, the stellar mass of MOSDEF galaxies correlates with their dynamical mass, thus stellar mass can be used as a rough proxy for the gravitational potential well. For simplicity, we have retained the factor of 2 in the denominator as the impact that feedback has on the ISM is likely sensitive to the entire galaxy mass, not just the mass contained within the half-light radius.

\subsection{Dynamical and Baryonic Masses}
\label{sec:alt_mass}

In addition to stellar masses, we also consider dynamical (\Mdyn), and baryonic (\Mbar\ = $M_{\star}$ + \Mgas) masses, as these masses may better trace the gravitational potential well of the galaxies. The procedure to calculate \Mdyn\ and \Mgas\ is described in \cite{Price20} and we refer readers there for more details. Briefly, \Mdyn\ was calculated as
\begin{equation}
	   M_{\rm{dyn}} = k_{\rm{tot}}(R_{\rm{E}}) \frac{V_{circ}(R_{\rm{E}})^2 R_{\rm{E}}}{G},
\end{equation}
where \RE\ is the effective radii, $k_{\rm{tot}}(R_{\rm{E}})$ is the virial coefficient and $G$ is the gravitational constant. For galaxies with resolved and detected rotation measured from 2D spectra, circular velocities can be calculated as: $V_{\rm{circ}}(R_{\rm{E}}) = \sqrt{V(R_{\rm{E}})^{2} + 3.35\sigma_{V,0}^{2}}$, where $\sigma_{V,0}$ is the intrinsic galaxy velocity dispersion \citep{Price20}. Otherwise, circular velocities are  calculated by assuming a fixed value of intrinsic rotation velocity divided by intrinsic galaxy velocity dispersion. Gas masses are estimated using the Kennicutt-Schmidt \citep{Kennicutt89} relation between $\rm{\Sigma_{SFR}}$ = SFR/(2$\pi R_{\rm{E}}^{2})$ and $\rm{\Sigma_{gas}}$ = $M_{\rm{gas}}$/(2$\pi R_{\rm{E}}^{2})$, where SFRs are  derived from \HA\ and \HB\ observations (if available) or SED fitting. In the LRIS-flow sample, 136 galaxies have measured dynamical and baryonic masses\footnote{Galaxies without a robustly measured \RE\ do not have a measured \Mdyn\ or \Mbar.}, with ranges of 9.1 < log($M_{\rm{dyn}}$/$M_{\odot}$) < 11.9 and 9.5 < log($M_{\rm{bar}}$/$M_{\odot}$) < 11.2, and medians log($M_{\rm{dyn}}$/$M_{\odot}$) = 10.3 and log($M_{\rm{bar}}$/$M_{\odot}$) = 10.4.

\subsection{Measurements}
\label{spec_indv}

\subsubsection{Outflow Velocity Measurements}
\label{sec:alt_vel}

Using systemic redshifts, LIS absorption line redshifts, and \LyA\ emission redshifts, we measure centroid outflow velocities from the redshift difference:

\begin{equation}
\Delta v_{\rm{LIS}} = \frac{c(z_{\rm{LIS}} - z_{\rm{sys}})}{1 + z_{\rm{sys}}}\quad 
{\rm{and}}\quad 
\Delta v_{\rm{Ly\alpha}} = \frac{c(z_{\rm{Ly\alpha}} - z_{\rm{sys}})}{1 + z_{\rm{sys}}},
\end{equation}
where $z_{\rm{sys}}$ is the systemic redshift from optical emission lines. In addition to centroid outflow velocities, another technique for estimating outflow velocity uses the blue wings of the absorption line profile. In general, \Vlis\ may include both outflowing gas and interstellar gas at or near $z_{\rm{sys}}$. The interstellar gas could then shift the line profile to lower velocities, so that the true outflowing gas is better traced by the blue wings of the absorption line profile.

To estimate the velocity of the blue wings, previous studies have either used the outflow velocity where the absorption feature reaches some percent of the continuum level \citep{Martin05, Weiner09, Chisholm15} or the maximum velocity where the absorption feature returns to the continuum level \citep{Steidel10, Kornei12, Rubin14, Prusinski21}. We consider both the outflow velocity at 80\% of the continuum (\Vfrc) and maximum outflow velocity (\Vmax) following a similar approach as \cite{Kornei12}. Using the normalised spectra, we identify the absolute minimum of a detected absorption feature, then move towards shorter wavelengths, checking the sum of the flux and its uncertainty at each wavelength step. We record the first wavelengths at which this sum exceeds 0.8 and 1.0, perturb the spectrum by its error spectrum, and repeat the same procedure many times. The average and standard deviation, after 3$\sigma$ clipping, of the trials are  then used to calculate \Vfrc, \Vmax, and their uncertainties. This process was repeated for each detected LIS feature, listed in Table \ref{tab:spectral_windows}, adopting \Vfrc\ and \Vmax\ as the average of the detected LIS features.  A similar bootstrap method for calculating uncertainties is applied for apparent optical depth (Section \ref{sec:depth}) and equivalent width (Section \ref{sec:EW}). For objects in the LRIS-flow sample, \Vfrc\ (\Vmax) ranges from $-$26 to $-$990 km s$^{-1}$ ($-$47 to $-$1090 km s$^{-1}$) with a median of $-$428$\pm$29 km s$^{-1}$ ($-$574$\pm$29 km s$^{-1}$).

\subsubsection{Apparent Optical Depth}
\label{sec:depth}

The LIS absorption lines analysed in this work are typically saturated, as observed in other studies of $z$ $\sim$ 2 -- 3 galaxies \citep{Shapley03, Trainor15, Du18}. As the LIS lines are saturated, the line depth provides a measure of the metal covering fraction along the line of sight, rather than the metal column density. Furthermore, with the LIS absorption lines falling on the flat part of the curve of growth, we cannot measure their optical depth. Instead, we measure the apparent optical depth (\depth) directly from the flux ratio:
\begin{equation}
	   \tau_{a} =  -\ln \left[\frac{F_\lambda}{F_{\lambda,\  \rm{cont}}} \right],
\end{equation}
where $F_{\lambda}$ is the observed flux and $F_{\lambda,\  \rm{cont}}$ is the continuum flux\footnote{We note that \depth\ is sensitive to the spectral resolution; i.e., a lower "optical depth" would be measured in a lower-resolution spectrum.}. For each detected LIS feature listed in Table \ref{tab:spectral_windows}, the flux at the rest wavelength is calculated by weighting the flux from the two nearest pixels. If the weighted flux is negative, then the \depth\ of that line is not calculated. The average \depth\ from individually-detected LIS lines is adopted as \depth\ for the galaxy.

\subsubsection{Equivalent Width}
\label{sec:EW}

The equivalent widths of the LIS features ($W_{\lambda}$) are measured by summing the normalised absorbed flux enclosed between the edges of a feature's spectral window (Table \ref{tab:spectral_windows}). The average equivalent width of individually-detected LIS lines is taken as \Wlis\, and its uncertainty is estimated by adding the 1$\sigma$ error bar of the detected LIS lines in quadrature. The equivalent width of \LyA\ (\Wlya) is measured following the procedures given in \cite{Kornei10} and \cite{Du18}. Throughout this work, $W_{\lambda}$ refers to the rest-frame value and is negative for absorption features. 

\subsection{Composite Spectra}

To evaluate the average outflow velocities in bins of other galaxy properties, we construct composite spectra by sorting the galaxies into equal-number bins according to various physical properties (e.g., SFR, mass, \SSFR, inclination, \depth, and $W_{\lambda}$). The composite spectrum is computed by shifting each galaxy’s blue and red spectra into the rest-frame, converting to luminosity density, interpolating onto a new wavelength grid, and taking the unweighted average of the spectra for all galaxies contributing to the composite. We refer readers to \cite{topping2020} and \cite{Reddy22} for more details on how the composite and associated error spectra were calculated\footnote{The code used to create the composite spectra is adapted from \cite{Shivaei18}; \url{https://github.com/IreneShivaei/specline/}}. Using the same techniques described for individual objects (Section \ref{spec_indv}), \Vfrc, \Vmax, \depth, and \Wlis\ were measured from the composite spectra. Centroid velocities (\Vlis\ and \Vlya) for the composite were measured in a similar way as \Vfrc\ and \Vmax\ (Section \ref{sec:alt_vel}), using the normalised spectra to measure the wavelength at the absolute minimum of the detected LIS absorption trough or at the maximum of the \LyA\ peak.

\section{Results}
\label{sec:4}

\begin{figure*}
  \includegraphics[height=4.2in, keepaspectratio]{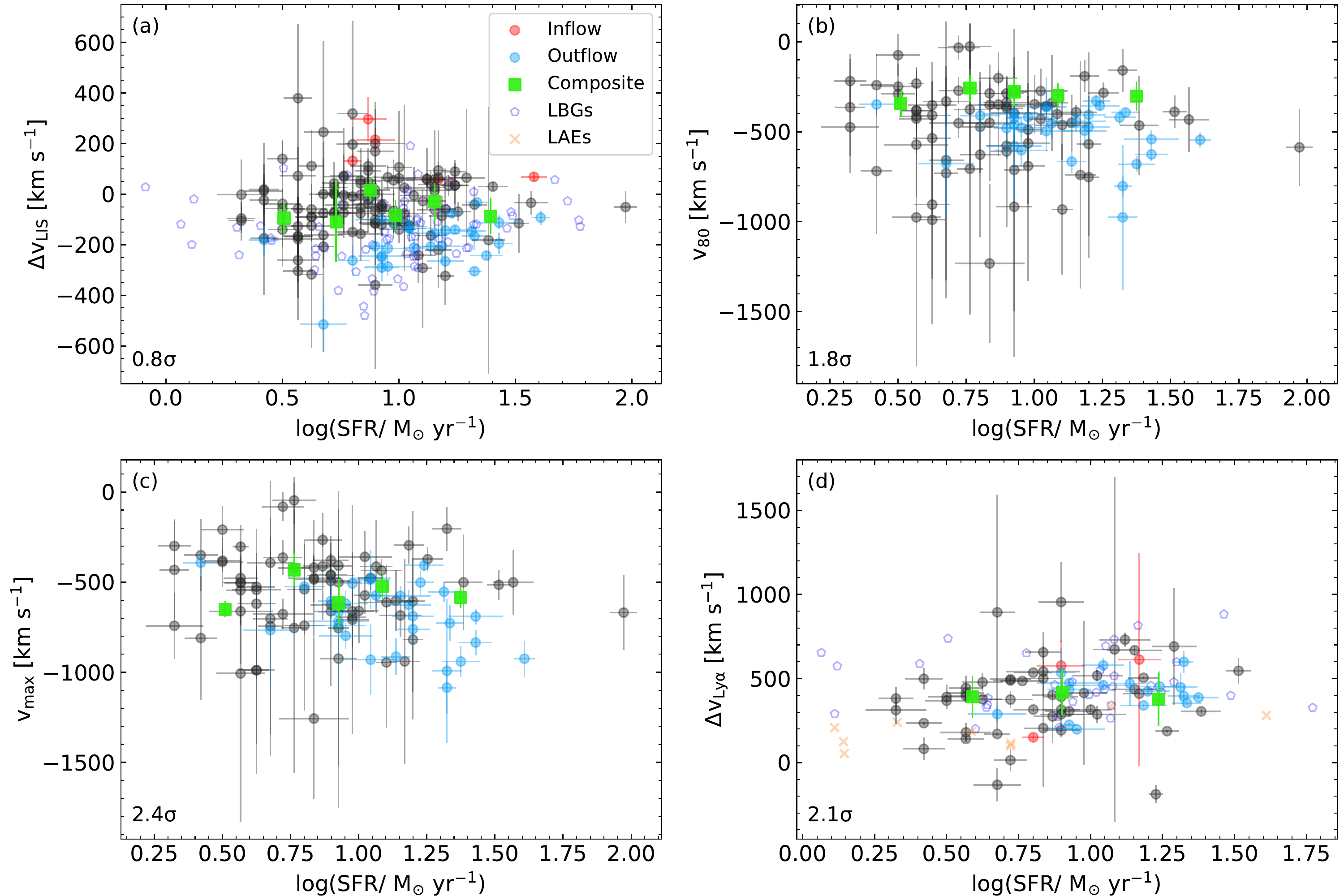}
  \vspace{-0.2cm}
  \caption{Outflow velocity versus log(SFR). Panel (a) \Vlis, Panel (b) \Vfrc, Panel (c) \Vmax, and Panel (d) \Vlya. Significant detections of outflows (\Vlis\ - 3$\sigma_{\Delta v_{\rm{LIS}}}$ < 0 km s$^{-1}$) or inflows (\Vlis\ - 3$\sigma_{\Delta v_{\rm{LIS}}}$ > 0 km s$^{-1}$) are shown as blue and red circles, respectively. Grey circles are galaxies with non-significant measured flows. Results from composite spectra, binning the galaxies by log(SFR), are shown as green squares. Blue pentagons are Lyman Break galaxies from \protect\cite{Erb06} and orange crosses are \LyA\ Emitter galaxies from \protect\cite{Erb16}. In the lower left corners, $\sigma$ is the number of standard deviations from the null hypothesis that the quantities are uncorrelated, based on a Spearman rank correlation test of the blue, red, and grey circles.
  \label{fig:vel-SFR}
  }
\end{figure*}

In this section, we present the relations between outflow velocity and several galactic properties. Table \ref{tbl:relation_results} summarises the results of Spearman correlation tests between \Vlis,\, \Vfrc,\, \Vmax,\, \Vlya,\, and the galaxy properties analysed in this work. Note that, when SFR[H$\rm{\alpha}$] is considered, 16 galaxies without robust measurements of the nebular dust attenuation, i.e. \HB\ detections, are not used in the correlation test.

\subsection{SFR and sSFR}

A key property is SFR, which sets the amount of mechanical energy and radiation pressure available in star-forming galaxies to drive outflows. Figure \ref{fig:vel-SFR} shows outflow velocity against SFR. We find that \Vmax\ and \Vlya\ are marginally correlated with SFR, such that higher SFR galaxies appear to have gas at larger velocities than lower SFR galaxies. While this trend is in agreement with the picture of galactic outflows driven by supernova or radiation pressure \citep{Chevalier85, Murray11}, it is surprising that these relations are found, given the small range of SFR probed by the LRIS-flow sample (SFR: 2 - 93 $M_{\odot}$ yr$^{-1}$). 

To increase our sample size, we include $z$ $\sim$ 2 Lyman Break galaxies (LBGs) and \LyA-emitters (LAEs) from the literature \citep{Erb06, Erb16}
\footnote{We recalculate the SFRs of the LBGs and LAEs using the same SED models as described in Section \ref{sec:SED}.}. Including LBGs in the Spearman test, the correlation between \Vlis\ and SFR decreases to 0.2$\sigma$. The lack of a correlation with \Vlis\ suggests, at a given SFR, that \Vlis\ may be biased to lower average outflow velocities possibly due to gas at rest near $z_{\rm{sys}}$. In this case, \Vfrc\ and \Vmax\ are likely more robust indicators of the outflow velocity (see Section \ref{sec:alt_vel}), thus a correlation is more likely to be seen over a smaller range in SFR. For \Vlya, the larger SFR range of LBGs increases the correlation with SFR slightly to 2.2$\sigma$, while the correlation increases to 3$\sigma$ when LAEs are included. However, the LAEs appear to have lower velocities at a given SFR, as seen in studies of LAEs \citep{Hashimoto13, Shibuya14}, complicating a direct comparison between these LAEs and the LRIS-flow sample.  
\begin{figure}
  \includegraphics[width=\columnwidth, keepaspectratio]{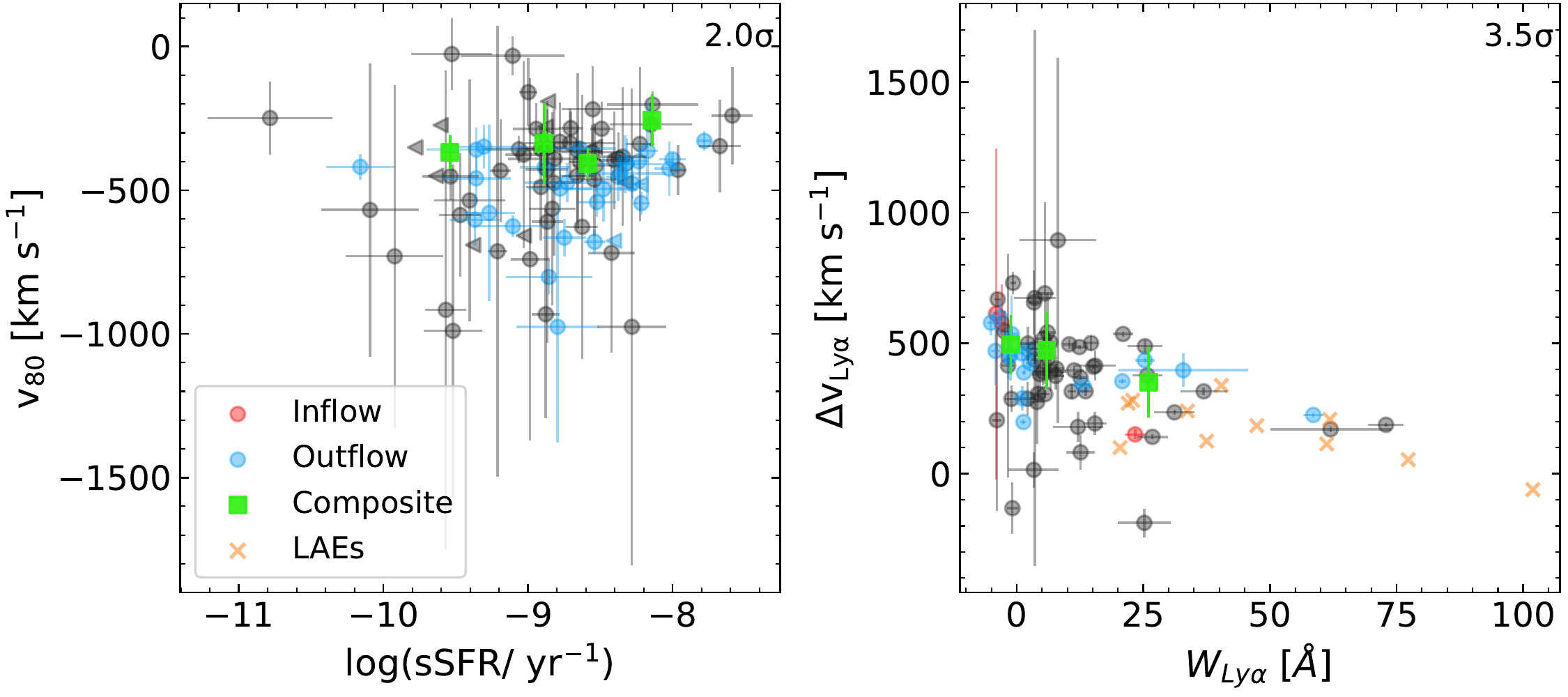}
  \vspace{-0.4cm}
  \caption{\textit{Left}: \Vfrc\ versus log(sSFR). \textit{Right}: \Vlya\ versus \Wlya. Triangles are upper limits for galaxies without H$_{\beta}$ detections. Same point and style as Figure \ref{fig:vel-SFR}.}
  \label{fig:sig-vel}
\end{figure}

Next, we investigate the dependence of outflow velocity on specific SFR. As sSFR is a tracer of both mechanical energy and gravitational potential energy (the latter due to the dependence of sSFR on \Mste), one might expect a correlation with outflow velocity. Of the four combinations of outflow velocities and sSFR, only a very marginal correlation is found between \Vfrc\ and sSFR at 2$\sigma$ (left panel of Figure \ref{fig:sig-vel}). These findings appear to contradict previous studies which have found no correlation between sSFR and outflow velocity \citep{Rubin10, Chisholm15, Prusinski21}. However, the finding of a marginal correlation with \Vfrc, but not \Vmax, is suspicious, as both should trace the high-velocity component of the outflow. If we restrict the correlation test to include only those galaxies for which the uncertainties in \Vlis\ imply that there are outflows with >3$\sigma$ significance, the correlation between \Vfrc\ and sSFR drops to 1.6$\sigma$. We conclude that the apparent marginal correlation between sSFR and \Vfrc\ is likely the result of galaxies with large uncertainties in \Vfrc\ (grey points), rather than having a physical origin.

\subsection{\textbf{\LyA} Equivalent Width}

The right panel of Figure \ref{fig:sig-vel} shows the relation between \Wlya\ and outflow velocity \Vlya. There is a clear anti-correlation between the two, such that objects with a larger \LyA\ equivalent width tend to have smaller outflow velocities. As shown in previous studies of LAEs at $z$ $\sim$ 2 -- 3 \citep{Erb14, Trainor15, Nakajima18}, this anti-correlation is likely tied to the column density of neutral hydrogen. For high column densities of gas at systemic redshift, \LyA\ photons resonantly scatter farther out in the wings to escape, increasing \Vlya\ while decreasing \Wlya. While we find a high significance of 3.7$\sigma$, this is weaker than the correlation reported by studies of LAEs. This is likely due to the smaller dynamic range in \Wlya\ covered in the LRIS-flow sample, where only 5\% of objects have \Wlya\ > 40 \AA\ compared to 42\% of LAEs in \cite{Erb16}. 
When LAEs are included in the Spearman test, the correlation between \Vlya\ and \Wlya\ increases, as expected, to 6.7$\sigma$.

\subsection{Inclination}
\label{sec:inc}

\begin{figure}
  \includegraphics[width=\columnwidth]{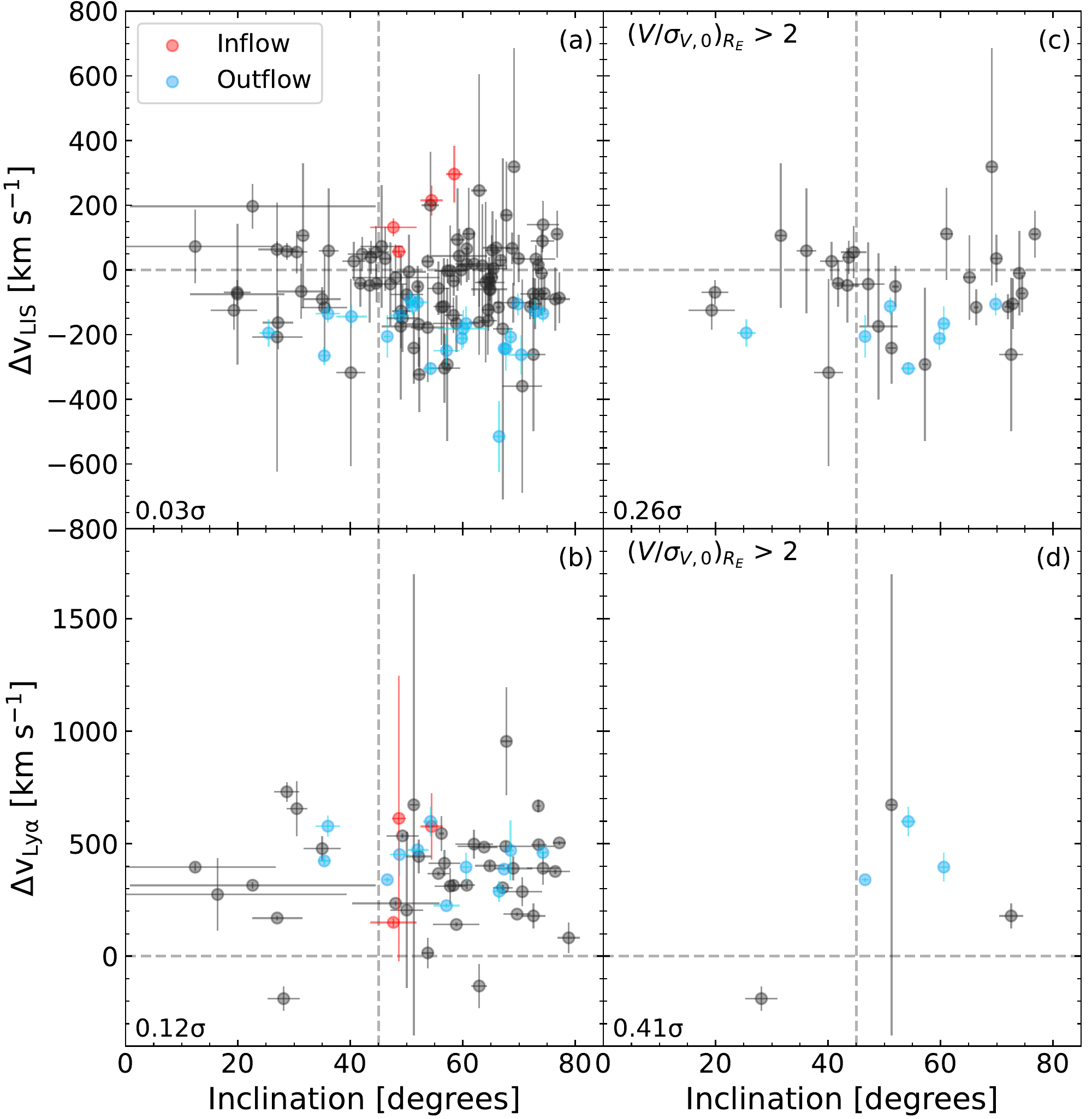}
  \vspace{-0.3cm}
  \caption{\textit{Top}: \Vlis\ versus inclination. \textit{Bottom}: \Vlya\ versus inclination. Panels (a) and (b) include all galaxies with a measured inclination, while panels (c) and (d) are limited to galaxies with $\left(V/\sigma_{V,0} \right)_{R_{E}}$ > 2. Same point and style as Fig. \ref{fig:vel-SFR}.
  }
  \label{fig:inc}
\end{figure}

In the canonical picture of galaxy flows, outflows emerge perpendicular to the disk in a biconical structure, while inflows occur along the major axis of the galaxy \citep{Heckman90, Katz93}. Within this picture, measured outflow velocities would strongly depend on inclination, with low-inclination (face-on) galaxies exhibiting faster outflows and weaker inflows compared to high-inclination (edge-on) galaxies. Studies of low-redshift galaxies have found a strong correlation of decreasing outflow velocity with increasing inclination, consistent with the physical picture described above \citep[e.g.,][]{Chen10, Concas19, Roberts19}. At higher redshifts ($z$ $\sim$ 1) , no strong correlation has been observed likely due to the difficulty of measuring inclination robustly for high-redshift galaxies with low-spatial-resolution observations and the lack of established disks. Despite these complications, studies have shown that low-inclination galaxies tend to exhibit outflowing gas, while inflowing gas is typically found in high-inclination ($i$ > 50) galaxies \citep{Kornei12, Rubin12, Rubin14}.

Here, we investigate the dependence of outflow velocity on galaxy inclination, where the inclination is calculated as the ratio of a galaxy’s semi-minor to semi-major axes, $i$ = cos($b$/$a$)$^{\rm{-1}}$. As shown in panels (a) and (b) of Figure \ref{fig:inc}, galaxies with significant inflows are only found at $i$ > 45$^{\circ}$, consistent with the physical picture of inflowing gas entering along the major axis of the galactic disk. For galaxies with significant outflows, 18/22 have inclinations above 45$^{\circ}$ (edge-on), which, at face value, is inconsistent with the canonical picture of bi-polar outflows\footnote{We note that the lack of galaxies with low inclinations is unlikely due to selection effects. In the parent MOSDEF sample, about 4\% of the sample has an inclination below 25$^{\circ}$, which is very similar to the 5\% of galaxies in the LRIS-flow sample.}. In part, the large fraction of outflowing galaxies in edge-on galaxies may reflect the lack of thin disks and/or the difficulty of measuring structural properties robustly for high redshift galaxies. Without established disks, the path with the lowest ambient gas pressure may not be along the minor axis, allowing outflows to escape at various angles, thus there would be no relation between inclination and \Vout. \cite{vanderWel12} found that the structural properties of galaxies in the CANDELS fields were accurate within $\sim$10$^{\circ}$ of the ’true’ properties using simulated galaxies images with known light distributions.

To explore this issue further, we investigate a rotation-dominated subsample, with $\left(V/\sigma_{V,0} \right)_{R_{E}}$ > 2 \citep{Price20}. This subsample should resemble local star-forming disks more closely than clumpy, irregular galaxies with large velocity dispersions, such that inclination may be better measured from the axis ratio. Panels (c) and (d) of Figure \ref{fig:inc} show outflow velocity against inclination for the subsample. Significant outflows are still primarily found in high inclination galaxies, with 6/7 of the galaxies inclined above 45$^{\circ}$. As with the full sample, there is no significant correlation between inclination and outflow velocity for the significantly outflowing galaxies. These results may suggest that the covering fraction of outflowing material is quite large, such that outflows are measurable even at high inclinations. Recently, \cite{Chen21} stacked \LyA\ spectral profiles of 59 star-forming galaxies at $z$ = 2 −- 3 galaxies and found an excess emission in the blueshifted component of \LyA\ along the minor axis, indicating a high covering fraction of outflowing gas.

\subsection{SFR Surface Density}
\label{sec:SSFR_results}

The connection between \SSFR\ and galactic outflows has been investigated in several studies \citep{Steidel10, Kornei12, Rubin14, Chisholm15, Davies19, Prusinski21}. Specifically, it has been suggested that regions with higher \SSFR, which traces the concentration of star formation in a galaxy, will be more efficient at injecting energy and momentum into the ISM from overlapping supernovae or stellar winds from massive stars, resulting in conditions amenable for launching outflows. Additionally, in this study, we consider a possible \Vout-\sSSFR\ relation. A correlation between \sSSFR\ and \Vout\ may be expected if galaxies with high \SSFR\ and low gravitational potential (or low stellar mass) are more efficient in launching outflows \citep{Reddy22}. We investigate both \SSFR\ and \sSSFR, and find no significant correlations (see Figure \ref{fig:all_SSFR}), which may be due to the limited dynamic range and/or the spatial resolution of our data (see Section \ref{sec:SSFR_sig}).

There is a debate in the literature about the existence of a \Vout$-$\SSFR\ relation. \cite{Kornei12} used a sample of  72 star-forming galaxies at $\rm{z\sim1}$ and found that galaxies with higher \SSFR\ had faster outflow velocities. However, \cite{Kornei12} proposed that such a relation is only present when inflowing and outflowing galaxies are considered together, with no significant trend between \SSFR\ and velocity among galaxies where the latter is negative (indicating outflows). \cite{Davies19} found that outflow velocity is related to \SSFR\ as \Vout\, $\propto \Sigma_{\rm{SFR}}^{0.34}$, using integral field unit observations of 28 star-forming galaxies at $\rm{z\sim2}$. The \cite{Davies19} study traced denser, ionised outflowing gas using the narrow and broad components of \HA\ emission, which may be partially broadened by shocks or turbulent mixing layers potentially making \HA\ less reliable for measuring outflow velocities. Both \cite{Steidel10} and \cite{Rubin14} found no correlation between outflow velocity with \SSFR. The limited dynamic range of SFR probed in these studies, similar to the range in our sample, is likely a contributing factor to their results.

\section{Discussion}
\label{sec:5}

\subsection{Outflow Driving Mechanisms}
\label{sec:driver}

The physical picture underlying observed trends among star-formation properties and outflow velocities have been considered in several theoretical and observational studies \citep{Chevalier85, Ferrara06, Steidel10, Murray11, Sharma12}. There are two commonly invoked mechanism for launching galactic-scale outflows in star-forming galaxies: (1) mechanical energy injected by supernovae \citep[“energy-driven”;][]{Chevalier85}; and (2) momentum injected by supernovae or radiation pressure from massive stars acting on dust grains \citep[“momentum-driven”;][]{Murray05, Murray11}. In the energy-driven case, mechanical energy from multiple, overlapping supernovae thermalises a large fraction of nearby gas into a hot over-pressured bubble. As the bubble expands adiabatically through the disk, it sweeps up ambient ISM material until it is ejected from the galaxy. Within the hot wind, ram pressure accelerates entrained cold gas clouds. Outflows driven by mechanical energy are predicted to scale weakly with star formation: \Vout\ $\propto \rm{SFR}^{0.2}$ \citep{Ferrara06} or \Vout\ $\propto \rm{SFR}^{0.25}$ \citep{Heckman00} and \Vout\ $\propto \Sigma_{\rm{SFR}}^{0.1}$ \citep{Chen10}. In the momentum-driven case, momentum is injected into the ISM by supernovae that accelerates cold gas, or radiation pressure from the absorption and scattering of photons on dust grains accelerating cold gas coupled to the dust. If the outflows are purely radiatively driven, the outflow velocity is predicted to scale strongly  with star formation activity: \Vout\ $\propto \rm{SFR}$  \citep{Sharma12} and \Vout\ $\propto \Sigma_{\rm{SFR}}^{2}$ \citep{Murray11}. As these mechanisms are likely dominate under different galactic conditions, outflows could be driven by a combination of mechanical energy and radiation pressure. The power law scaling between outflow velocity and star formation activity would then fall between the energy- and momentum-driven cases. \cite{Murray11} used 1-D models to investigate this case, and found that radiation pressure initially drives cold gas to about the scale height of the galaxy. After $\sim$3 -- 5 Myr, the lifetime of massive stars, supernovae begin to occur, and cold gas is then driven by radiation pressure and ram pressure to hundreds of kiloparsecs from the galaxy. 

Along with energy and momentum, cosmic rays produced by supernovae may drive large-scale galactic outflows in star-forming galaxies \citep[see discussions in][]{Heckman17, Zhang18}. As they diffuse out of the galaxy, cosmic rays scatter several times off of magnetic inhomogeneities in the ISM, transferring momentum to the surrounding gas. Based on the diffusion timescale of cosmic rays in the Milky Way, the total momentum deposited by cosmic rays is comparable to the momentum injected by radiation \citep{Zhang18}. However, despite the promising potential of cosmic rays to drive outflows, there are many open questions. Cosmic rays can be destroyed by scattering off of ISM gas, creating pions. If this destruction timescale is significantly shorter than the diffusion timescale, then the total cosmic ray momentum available to drive outflows would be severely limited. Additionally, the coupling between cosmic rays and multiphase gas is unclear, with some simulations finding that cosmic rays can decouple from cold gas clouds \citep{Everett11}.

There is tension between studies of low and high-redshift galaxies regarding the existence of a relation between SFR and outflow velocity. At low redshifts, \cite{Martin05} found that outflow velocity traced by the Na I absorption line scales as \Vout\ $\propto$ SFR$^{0.35}$, covering four orders-of-magnitude in SFR. \cite{Chen10} and \cite{Sugahara17} found no significant relation with the Na I centroid velocity over smaller ranges in SFR. However, \cite{Sugahara17} found a similar power-law scaling (\Vout\ $\propto$\ SFR$^{0.25}$) when outflow velocity is defined using the blue wings of the absorption profile, rather than centroid velocities. Studies at higher redshift are often limited to a smaller dynamic range of SFR and fail to find a significant correlations between SFR and centroid or maximum outflow velocities \citep{Steidel10, Law12, Kornei12, Rubin14}.

\begin{figure}
  \includegraphics[width=3in, keepaspectratio]{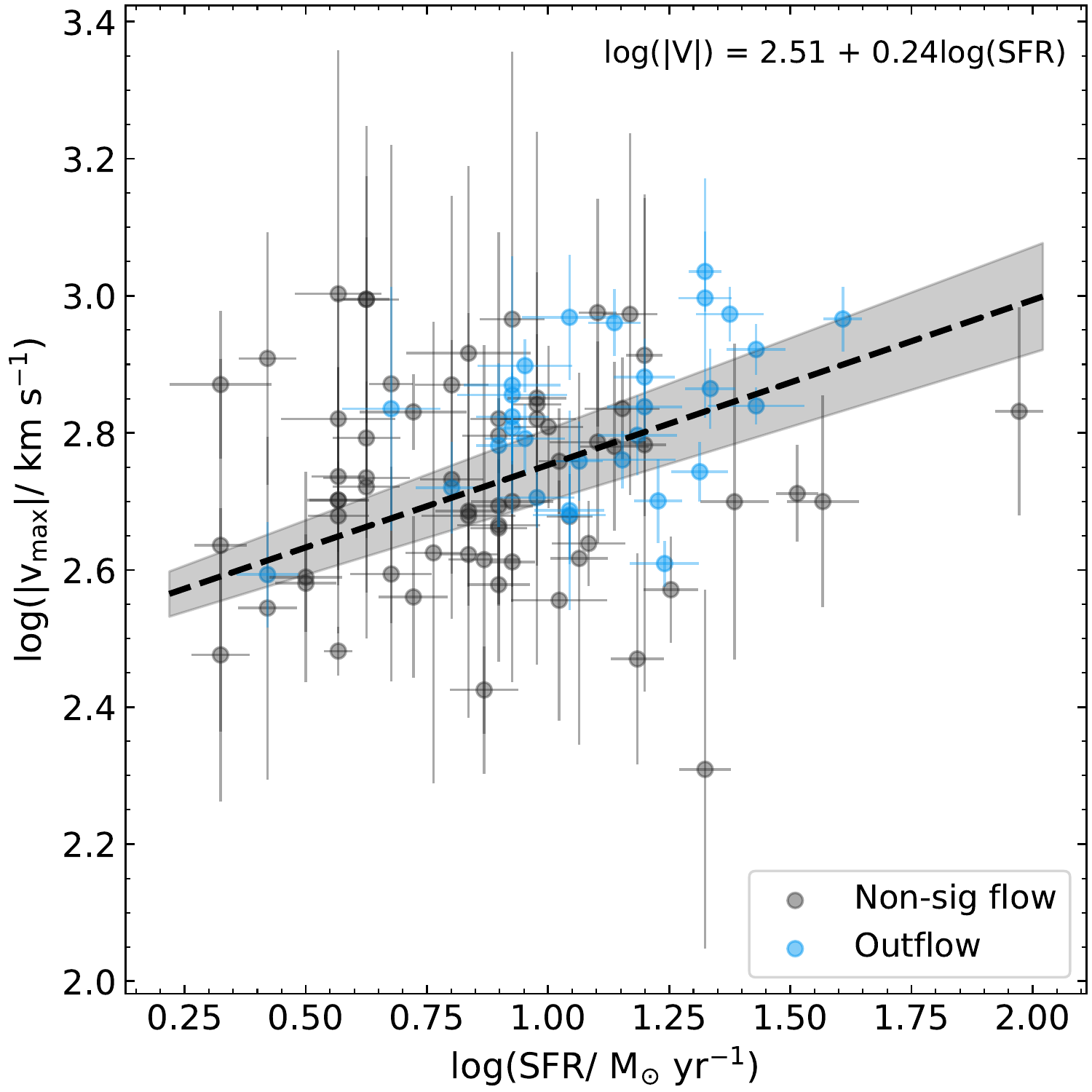}
  \vspace{-0.3cm}
  \caption{log(|\Vmax|) versus log(SFR). Blue circles are galaxies with 3$\sigma$-measured outflows. Grey circles are non-significant measured flows. The dashed black line and shaded region (68\% confidence intervals) is the best-fit line to the galaxies (blue and grey). The functional form of the line is listed in the upper-right corner.
  }
  \label{fig:SFR_power}
\end{figure}

Here, we investigate the marginally-correlated trend of \Vmax\ with SFR.
The power-law relations discussed above can be generalised as 
\begin{equation}
	   \rm{log}(|V|) = \alpha + \beta \rm{log}(SFR/ \rm{M_{\odot}}\ \rm{yr}^{-1}),
	   \label{equ:SFR_power}
\end{equation}
where V is the outflow velocity, $\alpha$ is the scaling factor, and $\beta$ is the power-law index. We adopt a Bayesian approach for calculating the linear regression to simultaneously fit possible combinations of $\alpha$ and $\beta$ to equation \ref{equ:SFR_power}, while accounting for the uncertainties in \Vmax\ and SFR. The results of the fitting are shown in Figure \ref{fig:SFR_power}. The best-fit power-law index is $\beta$ = 0.24$\pm$0.03. Our measurements of $\beta$ are consistent with the energy-driven case, suggesting that in these galaxies cool outflows are driven primarily from mechanical energy injected into the ISM from supernovae. However, we caution that $\beta$ is determined from a \textit{marginal correlation} between outflow velocity and SFR. In the next section, we explore the contribution of radiation pressure on outflow velocity.

\subsubsection{Radiation Pressure}
\label{sec:rad pressure}
\begin{figure}
  \includegraphics[width=\columnwidth]{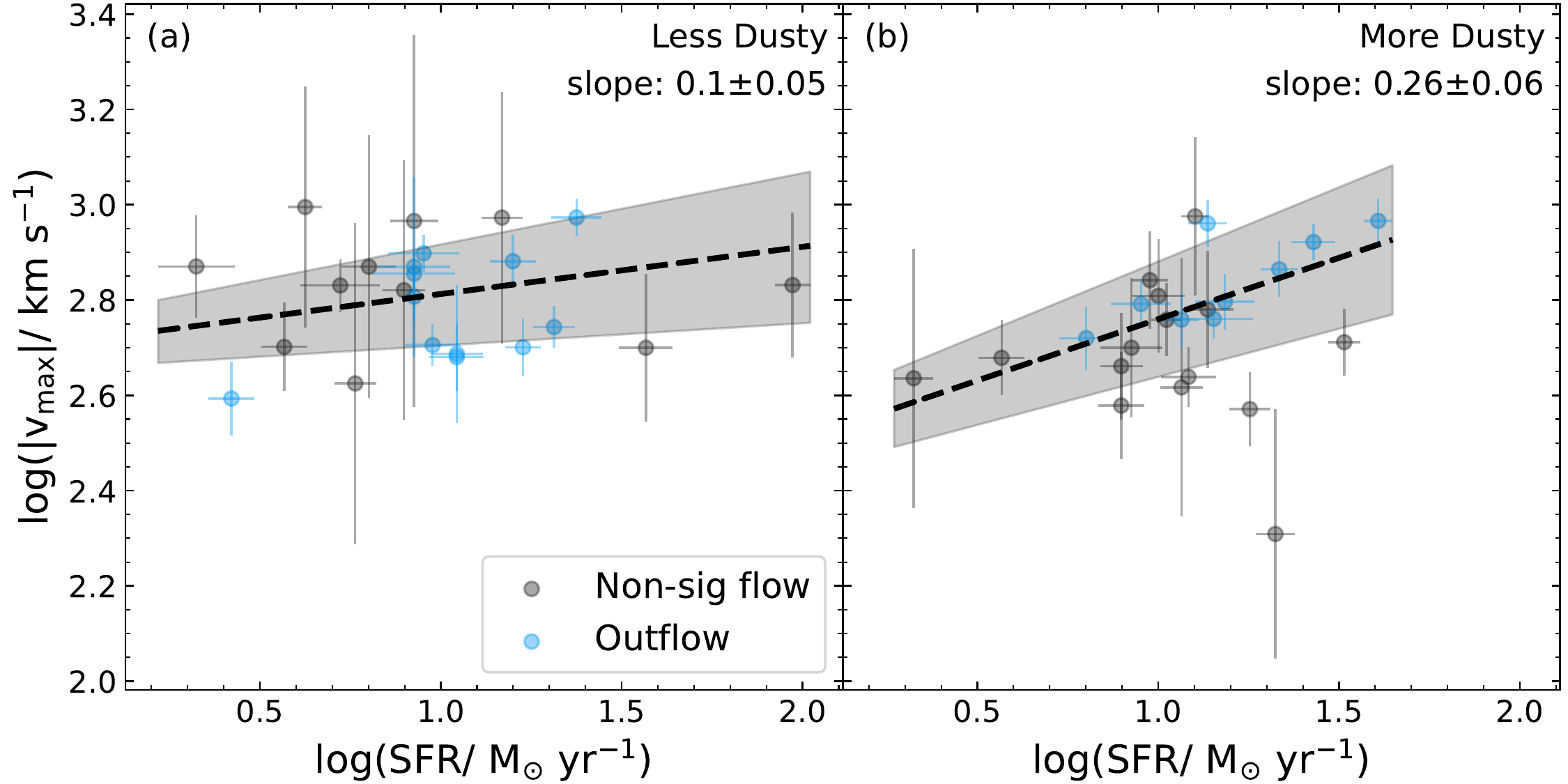}
  \vspace{-0.3cm}
  \caption{log(|\Vmax|) versus log(SFR). Panel (a) are "less dusty" galaxies ($\tau_{b}$ + $\sigma_{\tau_{b}}$ < $\tau_{b, \rm{median}}$), while panel (b) are "more dusty" ($\tau_{b}$ - $\sigma_{\tau_{b}}$ $\geq$ $\tau_{b, \rm{median}}$). Blue circles are galaxies with 3$\sigma$-measured outflows. Grey circles are non-significant measured flows. The dashed black line and shaded region (68\% confidence intervals) is the best-fit line to the sample. The slope and uncertainty of the line are listed in the upper-right corner.
  }
  \label{fig:dusty}
\end{figure}

While it appears that the outflows studied here are driven primarily by mechanical energy, these outflows may be driven by a combination of ram and radiation pressure. To explore the contribution of radiation pressure in driving outflows, we divided the sample into groups according to the dustiness of the galaxies. Dust is the cornerstone of momentum-driven outflows; without dust coupling to gas, radiation pressure on dust grains could not drive large amounts of gas out of galaxies \citep{Murray05}. If radiation pressure is negligible, then the slope of the \Vmax-SFR relation will remain consistent with the energy-driven case for galaxies with low and high dust content.

We parameterise the dustiness of galaxies using the Balmer decrement \citep{Calzetti94}:
\begin{equation}
    \tau_{b} = \rm{ln} \left(\frac{\rm{H\alpha/H\beta}}{2.86} \right),
\end{equation}
where \HA/\HB\ is the \HA\ to \HB\ line luminosity ratio and $\tau_{b}$, the Balmer decrement, is the difference in optical depths for \HB\ and \HA. The Balmer decrement is chosen over other dust metrics (e.g., E(B-V) or UV continuum slope) as it is sensitive to the reddening towards the ionised regions surrounding massive stars, which are more likely to have sufficient radiation pressure to drive outflows. We require that galaxies have 3$\sigma$ detections of \HA\ and \HB\ to calculate $\tau_{b}$. Galaxies with measured \HA/\HB\ < 2.86, the theoretical minimum value in the absence of dust for Case B recombination and T = 10000 K \citep{Osterbrock89}, are assigned $\tau_{b}$ = 0. Galaxies are divided into "less dusty" ($\tau_{b}$ + $\sigma_{\tau_{b}}$ < $\tau_{b, \rm{median}}$) and "more dusty" ($\tau_{b}$ - $\sigma_{\tau_{b}}$ > $\tau_{b, \rm{median}}$) groups\footnote{We note that the different ranges spanned by the "less dusty" and "more dusty" subsamples is primarily due to one outlier "less dusty" galaxy at log(SFR/$M_{\odot}$ yr$^{-1}$) = 1.9. After removing the outlier, the median SFR and interquartile range of the subsamples are similar, thus a comparison of the slopes is reasonable.}. Figure \ref{fig:dusty} shows that the slopes of the \Vmax-SFR relation between the two groups differ marginally (2$\sigma$), with "less dusty" galaxies having a weaker slope (0.10$\pm$0.05), while "more dusty" galaxies have a steeper slope (0.26$\pm$0.06). The difference in the slopes suggests that radiation pressure plays a minor role, along with ram pressure, in driving cool outflows. However, the dustiness traced by the Balmer decrement may not reflect the dust content of the outflows themselves, and without constraints on the other outflow phases (i.e., ionised and molecular), we cannot fully separate the contributions of ram and radiation pressure on galactic-scale outflows.

\subsubsection{Comparison to Simulations}

Lastly, it is useful to discuss the energy- vs. momentum-driven outflows in the context of simulations. A complete description of galactic-scale outflows and their impacts on galaxy evolution is challenging for simulations due to the different length scales of outflows. Both the ISM and large-scale galaxy features must be adequately resolved to capture relevant physical processes that generate outflows and how outflows interact with the CGM and IGM. These resolution requirements have led simulations of outflows to be performed in a relatively new generation of cosmological “zoom-in” simulations \citep{Hopkins14, Christensen16, Hopkins18}, where an individual galaxy is simulated to a high resolution within a larger, coarser cosmological volume. As simulations are not limited by observational constraints, one can directly probe the mass loading factor ($\eta$), the gas mass outflow rates normalised by SFR, and how it scales with the circular velocity (V$_{\rm{circ}}$) of the halo. In contrast to the \Vout-SFR relation, the $\eta-$V$_{\rm{circ}}$ relation is predicted to be steeper in the energy-driven case ($\eta \propto$ V$_{\rm{circ}}^{-\rm{2}}$) and shallower in the momentum-driven case ($\eta \propto$ V$_{\rm{circ}}^{-\rm{1}}$) \citep{Murray05}.

We have found that the outflows in the LRIS-flow sample are most consistent with an energy-driven scenario, which is supported by results from zoom-in simulations. \cite{Muratov15} used the Feedback in Realistic Environments \citep[FIRE-1;][]{Hopkins14} zoom-in simulations to analyse galactic-scale outflows and found a broken power law for the $\eta-$V$_{\rm{circ}}$ relation spanning the energy- and momentum-driven cases. \cite{Muratov15} concluded that the broken power law represented a transition from energy-driven outflows in dwarfs to momentum-driven outflows in higher mass halos. In a recent study, \cite{Pandya21} investigated outflows in the updated FIRE-2 simulations \citep{Hopkins18} and found that $z \sim$ 2 galaxies from low-mass dwarf halos to Milky Way-mass halos galaxies follow a $\eta-$V$_{\rm{circ}}^{-\rm{2}}$, in agreement with the energy-driven case. However, in both the FIRE-1 and FIRE-2 simulations, missing physics (e.g., radiation pressure from infrared multiple-scattering, type Ia SNe, cosmic rays, etc.) and the lower resolution of the ISM, compared to “resolved” ISM simulations, may lead to overestimated of mass-loading factors, thus radiation pressure could play a role in driving cool outflows.

\subsection{Significance of SFR Surface Densities on Outflows}
\label{sec:SSFR_sig}

As discussed in Section \ref{sec:SSFR_results}, outflow velocity does not appear to correlate significantly with the star-formation-rate surface density over the dynamic range of our sample. This result appears to be in tension with several other studies at low and intermediate redshifts, which find a weak \Vout-\SSFR\ relationship \citep{Chen10, Kornei12, Chisholm15}. However, there are three possible reasons for the lack of an observed relation: (1) there is ambiguity regarding the actual location of the gas and its coupling to the star formation activity, (2) \SSFR\ and outflow velocity may be correlated on spatial scales that are unresolved by the LRIS observations, and (3) the relationship between outflow velocity and \SSFR\ may be weak over the dynamic range of \SSFR\ probed in our sample. We discuss each of these possibilities below.

\subsubsection{Where is the Absorbing Gas?}
\label{sec:distance}

Using starlight from the galaxy as the background light source against which interstellar absorption and \LyA\ emission is measured (i.e., down-the-barrel observations) provides valuable information about gas flows. However, these observations tell us very little about the spatial location of the absorbing gas. To gain a better insight into the physical mechanisms that drive galactic outflows, and their effects on their host galaxies, we would require precise measurements of the physical location of the gas, so that position and velocity could be simultaneously constrained. 

The LRIS spectra, which probe down the barrel of the galaxy, only provide a surface-brightness weighted absorption profile for each observed LIS line, integrated along the entire line of sight. Although the different LIS lines appear to have similar profiles after integrating along the line of sight, there is no way of knowing where the bulk of the absorption is occurring relative to the galaxy. Furthermore, after integration, material from both past and current outflows affects the shape of the absorption line profiles. This indicates that the observed absorption may originate from regions far enough away from the galaxy that the outflow velocity traced by LIS absorption has uncoupled from changes in \SSFR. However, using a sample of close angular pairs of galaxies at $z$ $\sim$ 2 -- 3, \cite{Steidel10} found that LIS absorption line profiles are dominated by gas within $\sim$10 kpc of the galaxy. Based on $\left< \Delta v_{\rm{LIS}} \right>$ and $\left< v_{\rm{max}} \right>$ of the LRIS-flow sample, outflowing gas could exceed that distance in a roughly 50 Myr dynamical timescale, travelling 10 to 36 kpc. The observed absorption line profiles are likely originating in gas that is dynamically connected to recent star formation, thus the location ambiguity introduced by the down-the-barrel observations is unlikely the reason why we find no correlation between \Vout\ and \SSFR.

\subsubsection{Limited Spatial Resolution}
\label{sec:low_res}

Another possible explanation for the observed lack of correlation between outflow velocity and \SSFR\ is the limited spatial resolution probed by the LRIS spectroscopy. In particular, if the velocity of outflowing gas is coupled to \SSFR\ on scales smaller than a few kpc, then such a coupling may be masked by seeing-limited spectroscopy. Several studies have suggested that \SSFR\ and outflow velocity are correlated on small $\sim$kpc spatial scales. \cite{Bordoloi16} found that outflow velocities from individual star-forming knots in a lensed galaxy at $z$ $\sim$ 1.7 are correlated to the \SSFR\ of the knots, suggesting that outflows are ‘locally sourced’.
Similarly, \cite{Davies19} created high S/N stacks of IFU \HA\ observations from 28 $z$ $\sim$ 2.3 galaxies in bins of resolved physical properties. From their analysis, \cite{Davies19} concluded that \SSFR\ and outflows are closely related on 1 -- 2 kpc scales. In the LRIS-flow sample, $\sim$54\% of the galaxies have an effective radius, as measured from HST imaging, >2\,kpc, thus the seeing-limited observations could "wash-out" the small scale structure where \SSFR\ and outflows may be correlated.

\subsubsection{Strength of \Vout-\SSFR}
\label{sec:weak}

A final consideration is the predicted strength of the \Vout-\SSFR\ relation. As discussed in Section \ref{sec:driver}, outflow velocity is predicted to scale as \Vout\ $\propto \Sigma_{\rm{SFR}}^{0.1}$ in the energy-driven case up to \Vout\ $\propto \Sigma_{\rm{SFR}}^{2}$ in the momentum-driven case.

While we do not find a correlation between outflow velocity and \SSFR, here we investigate which case could be consistent with the observed \Vlis-\SSFR\ correlation. Two samples of outflowing galaxies (\Vlis\ < 0 km s$^{-1}$) are simulated following the predicted scaling relations of the energy- and momentum-driven cases over the dynamic range in \SSFR\ probed by the sample. Each value of \SSFR\ and \Vlis\ is perturbed assuming typical uncertainties of the measured values ($\sigma_{\Sigma_{\rm SFR}}$ = 0.1 $\rm{M_{\odot}}\ \rm{yr}^{-1}\ \rm{kpc}^{-2}$, $\sigma_{v_{\rm{LIS}}}$ = 100 km s$^{-1}$) and the intrinsic scatter in the observed \Vlis-\SSFR\ relation ($\sigma_{\rm{int}}$ = 0.09). This is repeated for 10,000 realisations in the energy- and momentum-driven cases. We find that none of the momentum-driven realisations yield a correlation as insignificant as the one that is observed, while 65\% of the energy-driven realisations are $\leq$1.77$\sigma$ correlated. These results suggest that the LRIS-flow sample more likely follows a weak relation as in the energy-driven case, rather than a steep relation predicted by the momentum-driven case.
As outflow velocity is likely only weakly dependent on \SSFR, a correlation between the two would require a large dynamic range in \SSFR\ to be observable. Our simulations imply that the small dynamic range of \SSFR\ of the LRIS-flow sample is likely responsible for the lack of an observed correlation between outflow velocity and \SSFR.

\subsubsection{\SSFR\ Threshold}
\label{sec:threshold}
\begin{figure}
  \includegraphics[width=3in, keepaspectratio]{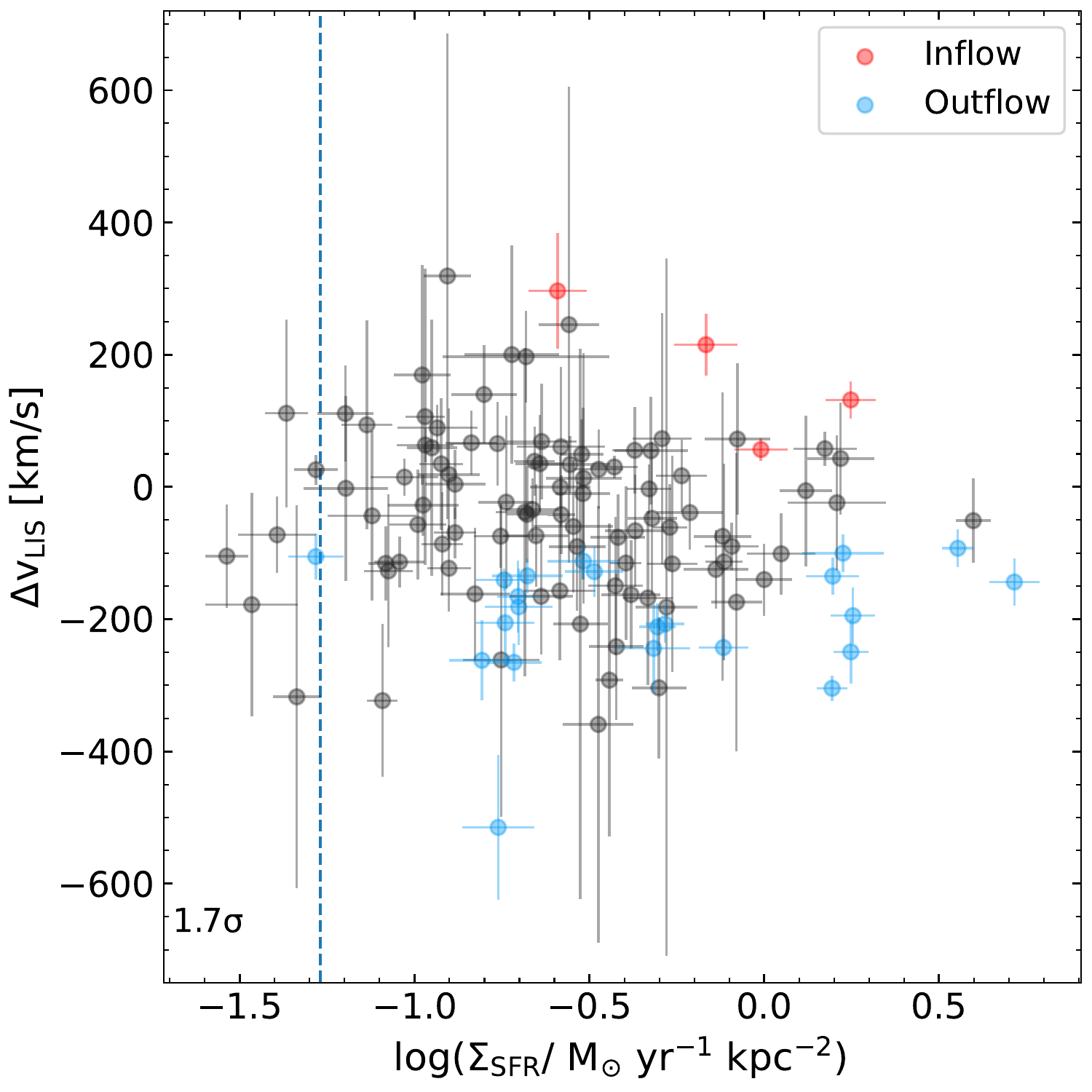}
  \vspace{-0.25cm}
  \caption{\Vlis\ vs log(\SSFR). Blue, red, and grey circles are galaxies with significant outflows, significant inflows, and non-significant flows, respectively. $\sigma$ is the number of standard deviations from the null hypothesis that the quantities are uncorrelated, based on a Spearman rank correlation test. The dashed line marks the threshold \SSFR\ from \protect\cite{Heckman02}.
  }
  \label{fig:SSFR_threshold}
\end{figure}

Starting with \cite{McKee77}, theoretical studies have long predicted that there exists a threshold \SSFR\ to launch galactic-scale outflows. Specifically, if the concentration of star formation is sufficiently high, then enough energy can be injected into surrounding gas allowing the gas to overcome its binding energy and escape the galaxy. Based on the \Vout-\SSFR\ relation observed in local starbursts galaxies, \cite{Heckman02} proposed a \SSFR\ threshold of $\sim$ 0.1 $\rm{M_{\odot}}\ \rm{yr}^{-1}\ \rm{kpc}^{-2}$.

We investigated if the LRIS-flow sample supports this threshold. The \cite{Heckman02} threshold is ${\sim} 0.05\ \rm{M_{\odot}}\ \rm{yr}^{-1}\ \rm{kpc}^{-2}$ for the \cite{Chabrier03} IMF assumed here. As shown in Figure \ref{fig:SSFR_threshold}, nearly every significant outflowing galaxy exceeds the \cite{Heckman02} threshold, with only one significant outflowing galaxy within the threshold given the measurement uncertainties. It is not surprising that our galaxies lie above this threshold, as the threshold itself is only approximate and the LRIS-flow sample does not probe to \SSFR\ significantly lower than the threshold. However, there is debate around the \cite{Heckman02} threshold, with some studies reporting galaxies with measurable outflow velocities down to $\sim$ 0.01 $\rm{M_{\odot}}\ \rm{yr}^{-1}\ \rm{kpc}^{-2}$ \citep{Rubin14, Chisholm15, Roberts20}.

\begin{figure}
  \includegraphics[width=\columnwidth, keepaspectratio]{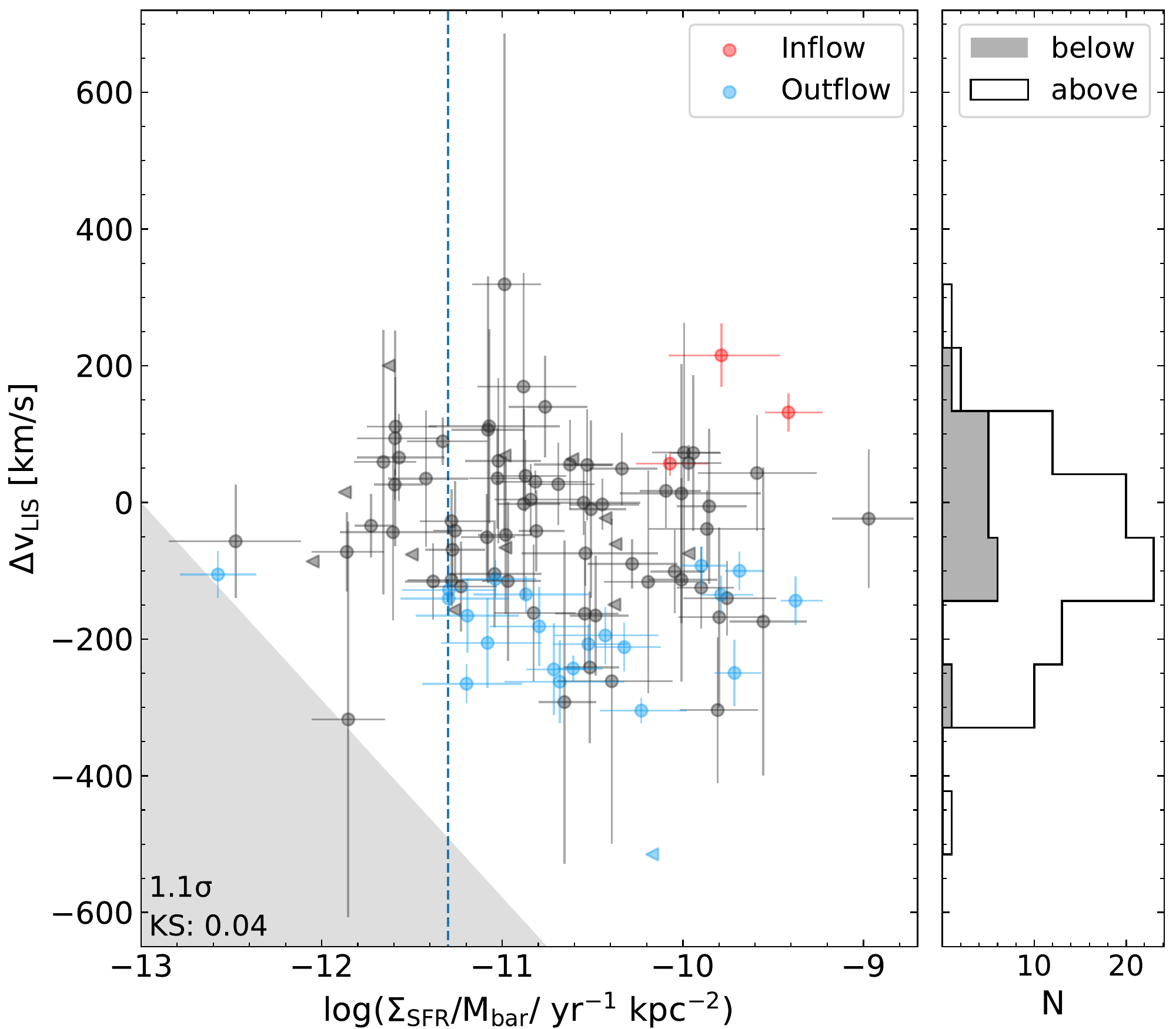}
  \vspace{-0.4cm}
  \caption{\textit{Left}: \Vlis\ vs log(\sSSFR). Blue, red, and grey circles are galaxies with significant outflows, significant inflows, and non-significant flows, respectively. $\sigma$ is the number of standard deviations from the null hypothesis that the quantities are uncorrelated, based on a Spearman rank correlation test. KS is the p-value of a Kolmogorov-Smirnov test (excluding significant inflowing galaxies) above and below log(\sSSFR/$\rm{yr}^{-1}\ \rm{kpc}^{-2}$) = $-$11.3, marked with a dashed line. \textit{Right}: Distribution of outflow velocities below (grey) and above (black outline) log(\sSSFR/$\rm{yr}^{-1}\ \rm{kpc}^{-2}$) = $-$11.3.
  }
  \label{fig:sSSFR_threshold}
\end{figure}

In addition to \SSFR, the mass of the galaxy, as a measure of the gravitational potential, may play an important role in launching outflows \citep[e.g.,][]{Reddy22}. In this case, one might observe faster outflows in galaxies with a high \SSFR\ and low potential (or mass). In Figure \ref{fig:sSSFR_threshold}, nearly all of the galaxies with significant outflows have a high \sSSFR, with only one found below log(\sSSFR/$\rm{yr}^{-1}\ \rm{kpc}^{-2}$) = $-$11.3. A Kolmogorov-Smirnov test (excluding the three galaxies with significant inflows) indicates a 4\% probability that galaxies below and above this threshold are drawn from the same parent distribution. Thus, outflows may only become common for galaxies with log(\sSSFR/$\rm{yr}^{-1}\ \rm{kpc}^{-2}$) > $-$11.3, while below this outflows tend to be weak\footnote{We note that the lack of galaxies in the lower left corner is unlikely due to selection effects. In the parent MOSDEF sample, about 8\% of the sample has a \sSSFR\ below log(\sSSFR/$\rm{yr}^{-1}\ \rm{kpc}^{-2}$) = $-$11.3 (using the same definition of \sSSFR\ given in equation \ref{equ:sSSFR}), which is very similar to the 7\% of galaxies in the LRIS-flow sample.}. Above log(\sSSFR/$\rm{yr}^{-1}\ \rm{kpc}^{-2}$) = $-$11.3, outflow velocity appears to uncouple from \sSSFR\, suggestive of a limit in the maximum allowable outflow speed, probably tied to the Eddington limit from radiation pressure on dust grains \citep{Murray05, Thompson05, Hopkins10}. This behaviour in \sSSFR\ is seen regardless of whether \SSFR\ is normalised by stellar, dynamical, or baryonic mass (see Figure \ref{fig:all_SSFR}).

\section{Conclusions}
\label{sec:con}

We use a sample of 155 typical star-forming galaxies at redshifts $z$ = 1.42 -- 3.48 to investigate how outflows vary with a number of galactic properties (e.g., SFR, mass, \SSFR, inclination, \sSSFR). The sample includes deep optical and FUV spectra obtained with the MOSFIRE and the Keck/LRIS spectrographs providing spectral covering of several LIS absorption lines and \LyA\ emission. The combination of MOSFIRE and LRIS spectra allow us to study outflows on an individual galaxy basis. Centroid velocities are measured from the redshift difference between $z_{\rm{sys}}$, $z_{\rm{LIS}}$, and/or $z_{\rm{Ly\alpha}}$, while fractional (\Vfrc) and maximum (\Vmax) outflow velocities are measured from the blue wings of LIS lines that may better trace outflowing gas. The galaxies exhibit blueshifted absorption features with a mean outflow velocities of \Vlis\ = $-$60$\pm$10 $\rm{km\ s^{-1}}$, \Vfrc\ = $-$468$\pm$29 $\rm{km\ s^{-1}}$, \Vmax\ = $-$591$\pm$29 $\rm{km\ s^{-1}}$, and redshifted \LyA\ emission with a mean velocity of \Vlya\ = 400$\pm$23 $\rm{km\ s^{-1}}$. We combined SFRs from SED modelling, \HA\ SFRs, and masses with galaxy areas based on effective radii to measure \SSFR\ and \sSSFR. Our main conclusions are as follows:
\begin{description}[leftmargin =1em]
    \setlength\itemsep{0.75em}
    \item[$\bullet$] We find marginal correlations between SFR and outflow velocities measured by \Vmax\ and \Vlya, such that higher SFR galaxies appear to have gas at larger velocities than lower SFR galaxies.
    \item[$\bullet$] Galaxies with significant outflows or inflows are found primarily at high inclinations ($i$ > 45$^{\circ}$). There appears to be no correlation between \Vout\ and inclination, which may be due to the difficulty of measuring inclination for these galaxies, or the lack of established disks (Section \ref{sec:inc}).
    \item[$\bullet$] Outflow velocity scales as \Vmax\ $\propto$ SFR$^{0.24\pm0.03}$. This scaling is in agreement with predictions for outflows driven by mechanical energy from supernovae, suggesting that supernovae are the primary driver of outflows in these $z$ $\sim$ 2 galaxies (Section \ref{sec:driver}). Radiation pressure acting on dusty material may play a minor role in the \Vmax-SFR relation (Section \ref{sec:rad pressure}).
    \item[$\bullet$] Outflow velocity is not correlated with \SSFR\ or \sSSFR, which may be due to limitations in the LRIS observations. After integrating along the entire line of sight, we lose vital spatial information about the absorbing gas. However, distant gas that has uncoupled from changes in \SSFR\ likely only provides a minor contribution to the measured outflow velocity (Section \ref{sec:distance}). The LRIS observations are not resolved, limiting our study to global \SSFR. Outflowing gas and \SSFR\ may be related on small spatial scales, such that any correlation between outflow velocity and global \SSFR\ disappears (Section \ref{sec:low_res}). Simulations suggest that the \Vout-\SSFR\ relation follows a weak scaling, as predicted by the energy-driven case, thus the small dynamic range of \SSFR\ probed by the LRIS-flow sample is likely the contributing factor for the absence of an observed correlation between \Vout\ and \SSFR\ (Section \ref{sec:weak}).
    \item[$\bullet$] Our sample agrees with the \cite{Heckman02} \SSFR\ threshold, and suggests a threshold in \sSSFR\ above which outflows are commonly detected. In the \Vout-\sSSFR\ relation, a KS-test indicates a 4\% probability that galaxies below and above log(\sSSFR/$\rm{yr}^{-1}\ \rm{kpc}^{-2}$) = $-$11.3 are drawn from the same parent distribution. Above this threshold, strong outflows are common, however, there appears to be a limit in the maximum allowable outflow speed, resulting in an insignificant correlation between outflow speed and \sSSFR\ above the aforementioned threshold (Section \ref{sec:threshold}).
\end{description}

Galactic-scale outflows are a critical component of the baryon cycle, influencing the environment and mass build-up of galaxies across cosmic time. Here, we have studied outflows in a large sample of $z$ $\sim$ 2 galaxies that push the limits of current ground-based facilities, with full night ($\sim$7.5 hrs) observations needed to obtain sufficiently high S/N spectra. We find that global galaxy properties and outflows are only weakly correlated and exhibit large scatter. However, outflows may be related to properties (e.g., \SSFR) across larger dynamical ranges or on smaller kpc scales. To build a better understanding of outflows, higher resolution spectroscopic data and spatially resolved imaging are necessary to constrain the geometry of outflowing gas. The increased sensitivity and field-of-view of the \textit{James Webb Space Telescope} and next generation of 30-m class telescopes will enable observations of galaxies with lower masses, lower star-formation rates, and higher star-formation-rate surface densities, allowing for studies of galactic outflows across orders of magnitude in galaxy properties.

\section*{Acknowledgements}


We acknowledge support from NSF AAG grants AST1312780, 1312547, 1312764, and 1313171, grant AR13907 from the Space Telescope Science Institute, and grant NNX16AF54G from the NASA ADAP program. We thank the 3D-HST Collaboration, which provided the spectroscopic and photometric catalogs used to select the MOSDEF targets and derive stellar population parameters. This research made use of Astropy,\footnote{\url{http://www.astropy.org}} a community-developed core Python package for Astronomy \citep{Astropy13, Astropy18}. We wish to extend special thanks to those of Hawaiian ancestry on whose sacred mountain we are privileged to be guests. Without their generous hospitality, most of the observations presented herein would not have been possible.

\section*{Data Availability}


The MOSDEF data used in this article is publicly available and can be obtained at \url{http://mosdef.astro.berkeley.edu/for-scientists/data-releases/}. The LRIS data used in this article is available upon request.



\bibliographystyle{mnras}
\bibliography{ref} 




\appendix

\section{}
\begin{figure*}
  \includegraphics[width=\linewidth, keepaspectratio]{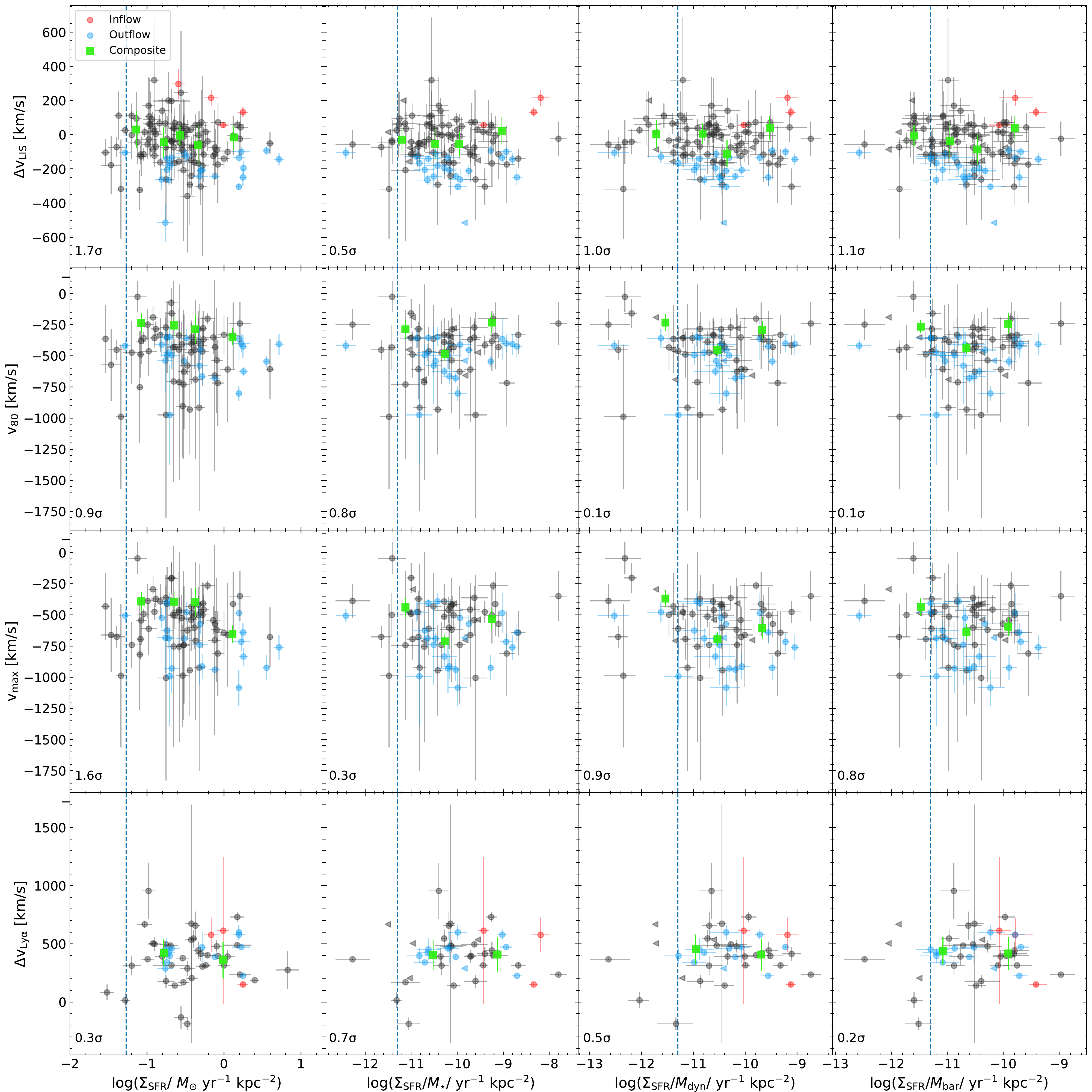}
  \vspace{-0.5cm}
  \caption{
  Plots of outflow velocity versus \SSFR\ and \sSSFR. Blue, red, and grey circles represent galaxies with significant (\Vlis\ - 3$\sigma_{\Delta v_{\rm{LIS}}}$ < 0 km s$^{-1}$) outflows, significant (\Vlis\ - 3$\sigma_{\Delta v_{\rm{LIS}}}$ > 0 km s$^{-1}$) inflows, and non-significant flows, respectively. Triangles (in \sSSFR\ panels) are upper limits for galaxies without H$_{\beta}$ detections. Results from composite spectra are shown as green squares. In the lower left corners, $\sigma$ is the number of standard deviations from the null hypothesis that the quantities are uncorrelated, based on a Spearman rank correlation test. For \SSFR, the dashed line marks the threshold \SSFR\ from \protect\cite{Heckman02}. For \sSSFR, the dashed line marks our proposed \sSSFR\ threshold; log(\sSSFR/$\rm{yr}^{-1}\ \rm{kpc}^{-2}$) = $-$11.3 (Section \ref{sec:threshold}).
  }
  \label{fig:all_SSFR}
\end{figure*}

Here we provide a table of statistical tests between \Vout\ and the galaxy properties analysed in this work (Table \ref{tbl:relation_results}), and plots of \Vout\ versus \SSFR\ and \sSSFR\ (Figure \ref{fig:all_SSFR}).
\clearpage

\begin{table}
  \caption{}
    \begin{tabular}{llccccc}
        \hline
        \hline
        Attribute  & Quantity & N & $\rm{\rho_{s}}$ & $\rm{\sigma}$ & $\rm{P_{s}}$ & KS\\
        (1) & (2) & (3) & (4) & (5) & (6) & (7)\\\hline
        
        \Vlis 
              & log(SFR[SED])                                & 147 & -0.07 & 0.81 & 0.42 & 0.33\\
              & log(sSFR[H$\alpha$])                         & 112 &  0.08 & 0.89 & 0.38 & 0.98\\
              & log($\rm M_{\star}$)                         & 147 & -0.07 & 0.78 & 0.43 & 0.94\\
              & log($\rm M_{dyn}$)                           & 129 &  0.07 & 0.76 & 0.45 & 0.12\\
              & log($\rm M_{bar}$)                           & 129 &  0.04 & 0.42 & 0.68 & 0.88\\
              & log($\Sigma_{\rm SFR[SED]}$)                 & 119 & -0.16 & 1.70 & 0.09 & 0.15\\
              & log($\Sigma_{\rm SFR[H\alpha]}/\rm M_{\star}$) & 91  & -0.05 & 0.49 & 0.62 & 0.67\\
              & log($\Sigma_{\rm SFR[H\alpha]}/\rm M_{dyn}$) & 90  & -0.10 & 0.95 & 0.34 & 0.33\\
              & log($\Sigma_{\rm SFR[H\alpha]}/\rm M_{bar}$) & 90  & -0.12 & 1.13 & 0.26 & 0.22\\
              & $i$                                          & 119 & 0.00 & 0.03 & 0.97 & 0.68\\
              & \Wlis\                                       & 112 & 0.00 & 0.01 & 0.99 & 0.62\\
              & $\tau_{\rm LIS}$                             & 145 & 0.07 & 0.86 & 0.39 & 0.27\\
        \hline
        \Vfrc 
              & log(SFR[SED])                                & 100 & -0.18 & 1.84 & 0.07 & 0.19\\
              & log(sSFR[H$\alpha$])                         & 77  &  0.23 & 2.04 & 0.04 & 0.04\\
              & log($\rm M_{\star}$)                         & 100 & -0.20 & 1.90 & 0.06 & 0.40\\
              & log($\rm M_{dyn}$)                           & 88  & -0.07 & 0.67 & 0.51 & 0.94\\
              & log($\rm M_{bar}$)                           & 88  & -0.03 & 0.31 & 0.75 & 0.81\\
              & log($\Sigma_{\rm SFR[SED]}$)                 & 79  & -0.10 & 0.90 & 0.37 & 0.72\\
              & log($\Sigma_{\rm SFR[H\alpha]}/\rm M_{\star}$) & 61  &  0.10 & 0.76 & 0.45 & 0.39\\
              & log($\Sigma_{\rm SFR[H\alpha]}/\rm M_{dyn}$) & 60  & -0.01 & 0.08 & 0.94 & 0.60\\
              & log($\Sigma_{\rm SFR[H\alpha]}/\rm M_{bar}$) & 60  &  0.02 & 0.12 & 0.90 & 0.96\\
              & $i$                                          & 79  &  0.02 & 0.14 & 0.89 & 0.81\\
              & \Wlis\                                       & 75  &  0.23 & 1.97 & 0.05 & 0.08\\
              & $\tau_{\rm LIS}$                             & 100 &  0.30 & 3.10 & 0.00 & 0.18\\
        \hline
        \Vmax 
              & log(SFR[SED])                                & 100 & -0.25 & 2.50 & 0.01 & 0.08\\
              & log(sSFR[H$\alpha$])                         & 77  &  0.08 & 0.65 & 0.52 & 0.24\\
              & log($\rm M_{\star}$)                         & 100 & -0.12 & 1.23 & 0.22 & 0.55\\
              & log($\rm M_{dyn}$)                           & 88  & -0.01 & 0.09 & 0.93 & 0.94\\
              & log($\rm M_{bar}$)                           & 88  & -0.04 & 0.37 & 0.71 & 0.84\\
              & log($\Sigma_{\rm SFR[SED]}$)                 & 79  & -0.18 & 1.62 & 0.11 & 0.33\\
              & log($\Sigma_{\rm SFR[H\alpha]}/\rm M_{\star}$) & 61  & -0.04 & 0.30 & 0.76 & 0.59\\
              & log($\Sigma_{\rm SFR[H\alpha]}/\rm M_{dyn}$) & 60  & -0.12 & 0.94 & 0.35 & 0.59\\
              & log($\Sigma_{\rm SFR[H\alpha]}/\rm M_{bar}$) & 60  & -0.10 & 0.76 & 0.45 & 0.96\\
              & $i$                                          & 79  &  0.03 & 0.25 & 0.80 & 0.42\\
              & \Wlis\                                       & 75  &  0.14 & 1.20 & 0.23 & 0.35\\
              & $\tau_{\rm LIS}$                             & 100 &  0.33 & 3.49 & 0.00 & 0.07\\
    \hline
        
    \end{tabular} 
  \label{tbl:relation_results}
\end{table}
\begin{table}
  \label{tbl:rel_sig}
  \contcaption{}
    \begin{tabular}{llccccc}
        \hline
        \hline
        Attribute  & Quantity & N & $\rm{\rho_{s}}$ & $\rm{\sigma}$ & $\rm{P_{s}}$ & KS\\
        (1) & (2) & (3) & (4) & (5) & (6) & (7)\\\hline
        
        \Vlya 
              & log(SFR[SED])                                & 71 &  0.25 & 2.15 & 0.04 & 0.20\\
              & log(sSFR[H$\alpha$])                         & 52 &  0.12 & 0.85 & 0.40 & 0.31\\
              & log($\rm M_{\star}$)                         & 71 &  0.11 & 0.90 & 0.37 & 0.41\\
              & log($\rm M_{dyn}$)                           & 62 &  0.11 & 0.82 & 0.41 & 0.08\\
              & log($\rm M_{bar}$)                           & 62 &  0.22 & 1.76 & 0.06 & 0.15\\
              & log($\Sigma_{\rm SFR[SED]}$)                 & 53 &  0.05 & 0.34 & 0.74 & 0.43\\
              & log($\Sigma_{\rm SFR[H\alpha]}/\rm M_{\star}$) & 38 &  0.11 & 0.67 & 0.51 & 0.98\\
              & log($\Sigma_{\rm SFR[H\alpha]}/\rm M_{dyn}$) & 37 &  0.08 & 0.50 & 0.62 & 0.92\\
              & log($\Sigma_{\rm SFR[H\alpha]}/\rm M_{bar}$) & 37 &  0.04 & 0.21 & 0.84 & 0.92\\
              & $i$                                          & 53 & -0.02 & 0.12 & 0.90 & 0.26\\
              & \Wlya\                                       & 69 & -0.40 & 3.53 & 0.00 & 0.03\\

        \hline
        $W_{\rm{LIS}}$ 
                        & log(SFR[SED])                                & 112 & -0.05 & 0.57 & 0.57 & 0.47\\
                        & log(sSFR[H$\alpha$])                         & 88  & -0.03 & 0.28 & 0.78 & 0.32\\
                        & log($\Sigma_{\rm SFR[SED]}$)                 & 90  & -0.07 & 0.70 & 0.49 & 0.48\\
                        & log($\Sigma_{\rm SFR[H\alpha]}/\rm M_{\star}$) & 70  & -0.05 & 0.38 & 0.71 & 0.32\\
                        & log($\Sigma_{\rm SFR[H\alpha]}/\rm M_{dyn}$) & 70  & -0.08 & 0.67 & 0.50 & 0.20\\
                        & log($\Sigma_{\rm SFR[H\alpha]}/\rm M_{bar}$) & 70  & -0.08 & 0.62 & 0.54 & 0.12\\
        \hline
        $W_{\rm{Ly}\alpha}$ 
                             & log(SFR[SED])                                & 69 & -0.30 & 2.59 & 0.01 & 0.03\\
                             & log(sSFR[H$\alpha$])                         & 50 & -0.03 & 0.19 & 0.85 & 0.99\\
                             & log($\Sigma_{\rm SFR[SED]}$)                 & 51 & 0.01 & 0.09 & 0.93 & 0.70\\
                             & log($\Sigma_{\rm SFR[H\alpha]}/\rm M_{\star}$) & 36 & -0.04 & 0.24 & 0.81 & 0.78\\
                             & log($\Sigma_{\rm SFR[H\alpha]}/\rm M_{dyn}$) & 35 &  0.00 & 0.03 & 0.98 & 0.88\\
                             & log($\Sigma_{\rm SFR[H\alpha]}/\rm M_{bar}$) & 35 &  0.07 & 0.39 & 0.70 & 0.88\\
        \hline
        $\tau_{\rm{LIS}}$   
                             & log(SFR[SED])                                & 145 &  0.10 & 1.24 & 0.22 & 0.04\\
                             & log(sSFR[H$\alpha$])                         & 110 &  0.03 & 0.27 & 0.79 & 0.46\\
                             & log($\Sigma_{\rm SFR[SED]}$)                 & 118 &  0.06 & 0.62 & 0.54 & 0.65\\
                             & log($\Sigma_{\rm SFR[H\alpha]}/\rm M_{\star}$) & 90  &  0.04 & 0.39 & 0.70 & 0.48\\
                             & log($\Sigma_{\rm SFR[H\alpha]}/\rm M_{dyn}$)  & 89  &  0.10 & 0.99 & 0.34 & 0.49\\
                             & log($\Sigma_{\rm SFR[H\alpha]}/\rm M_{bar}$) & 89  &  0.02 & 0.22 & 0.83 & 0.54\\
        \hline
                  
    \end{tabular}
    \begin{tablenotes}
        \item (1): Attribute is galactic property on the y-axis. 
        \item (2): Quantity is galactic property on the x-axis. 
        \item (3): Number of galaxies used to evaluate the correlation. 
        \item (4): Spearman rank correlation coefficient. 
        \item (5): Number of standard deviations by which the correlation deviates from the null hypothesis of no correlation. 
        \item (6): Spearman p-value. 
        \item (7): Kolmogorov–Smirnov test p-value from dividing the galaxies into two groups, above and below the median Quantity, and testing against the Attribute.
    \end{tablenotes}
    
\end{table}

\bsp	
\label{lastpage}
\end{document}